\documentclass[apj,twocolumn]{openjournal}

\usepackage{newtxtext,newtxmath,enumitem,multirow,ctable,CJK}
\usepackage{comment}
\usepackage[T1]{fontenc}
\usepackage[breaklinks,color links,citecolor=blue,urlcolor=blue]{hyperref}

\newcommand{\dun}{{\rm cm^2 \, s^{-1}}}

\newcommand{\orcidauthor}[3]{\author{\href{http://orcid.org/#1}{#2 \openin1 Orcid-ID.png \ifeof1 \else \hskip2pt\includegraphics[width=9pt]{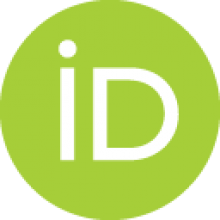}\fi}$^{#3}$}}

\graphicspath{{./}{figures/}}

\shorttitle{Cosmic Ray Feedback in Massive Halos}
\shortauthors{Quataert \& Hopkins}

\begin{document}

\title{\vspace{-0.8cm}Cosmic Ray Feedback in Massive Halos:  Implications for the  Distribution of Baryons\vspace{-1.5cm}}

\affiliation{$^{1}$Department of Astrophysical Sciences, Princeton University, Princeton, NJ 08544, USA}
\affiliation{$^{2}$TAPIR, Mailcode 350-17, California Institute of Technology, Pasadena, CA 91125, USA}

\orcidauthor{0000-0001-9185-5044}{Eliot Quataert}{1 *}
\orcidauthor{0000-0003-3729-1684}{Philip F. Hopkins}{2 **}

\email{* quataert@princeton.edu}
\email{** phopkins@caltech.edu}

\begin{abstract}

We use order of magnitude estimates and observational constraints to argue that feedback from relativistic cosmic rays (CRs) produced by massive black holes is likely to have a particularly large effect at radii of order the virial radius and larger in group-mass halos. We show that for a range of plausible (but uncertain) CR transport parameters and energetics, the pressure produced by CRs generated by the central massive black hole over its lifetime can be of order the thermal gas pressure in the outskirts of $\sim 10^{13-14} M_\odot$ halos (but not in more massive clusters). The properties of this CR feedback at low redshift are not well predicted by the radiative cooling rate of hot gas at smaller radii, which is often used as a proxy for `current' black hole feedback.  This is because most black hole growth happens early in massive halos, and CR transport timescales in halo outskirts are Gyr or more; the accumulated CR energy thus depends on the full history of black hole activity in the halo.  The large CR pressure in  group-mass systems  likely leads to CR-driven outflows that move gas from large halo radii to outside the virial radius. Such feedback would not be captured by current cosmological simulations that focus on mechanical black hole feedback; in particular, CR feedback  remains active even long after the mechanical feedback sourcing the CRs has turned off.  We speculate that this CR feedback may be important for explaining the weak lensing $S_8$ tension and the evidence for strong feedback at large halo radii from kinetic Sunyaev-Zeldovich measurements.  Prospects for testing this mechanism observationally and implementing the necessary physics in cosmological simulations are discussed.
\end{abstract}


\section{Introduction} \label{sec:intro}

The absence of ongoing star formation or a large reservoir of cool gas in galaxy groups and clusters requires a heating mechanism to balance the  cooling of the hot ($T\sim10^{7-8}~\mathrm{K}$) X-ray emitting plasma. Stellar feedback is energetically insufficient, so feedback from massive black holes (BHs) is generally believed to be the primary heating mechanism balancing radiative losses in massive halos \citep{Birzan2004,Croton2006,Fabian2012}.  In the prevailing model favored theoretically and observationally,  
jets from active galactic nuclei (AGN) heat the ambient halo gas.   Exactly how this coupling occurs is still not fully understood, but it is likely that a combination of mechanical feedback (e.g., bubbles, turbulence, waves generated by the jet) and/or relativistic particles (cosmic rays [CRs]) created by the jet are responsible (e.g., \citealt{Guo2008,Li2014}).

The physical picture of radiative cooling  regulated by jet feedback in massive halos is one developed primarily in the context of observations of the {inner regions of galaxy clusters in massive $\sim 10^{14-15} M_\odot$ halos}, where radiative cooling times are the shortest and where X-ray observations are the most diagnostic.  More recently, however, there are separate indications of a potentially important source of feedback in the {outer regions of lower halo mass $\sim 10^{13-14} M_\odot$ galaxy groups.}  This includes the tension between $S_8 = \sigma_8 (\Omega_m/0.3)^{1/2}$ measured by weak lensing vs the CMB \citep{Asgari2021,Amon2022b}.   This discrepancy can be reconciled if feedback in the outskirts of group-mass halos is significantly stronger than in most cosmological simulations \citep{Amon2022,Preston2023}; this can suppress the matter power spectrum near and exterior to the virial radius,  thus suppressing the weak lensing signal.  Separately, observations of the kinetic Sunyaev-Zeldovich effect, which are sensitive to the density of ionized lower temperature gas in halo outskirts, shows that the gas is more diffuse and radially extended than the dark matter to high statistical significance \citep{Boryana2024}.  This again points to strong feedback in halo outskirts \citep{Bigwood2024}.   Observations of the dispersion measures of fast radio bursts also find that most of the baryons do not reside in halos \citep{Connor2024}, though this is restricted to $\lesssim 5 \times 10^{12} M_\odot$ halos while the weak lensing and kSZ measurements are dominated by more massive halos.   And thermal Sunyaev-Zeldovich measurements cross-correlated with DES weak lensing show that $\simeq 10^{13-14} M_\odot$ halos have a total thermal energy less than the predictions of cosmological simulations with typical supernova and BH feedback \citep{Sandey2022}.

Cosmological simulations that reproduce X-ray observations of the gas fractions of groups and clusters (e.g., \citealt{vanDaalen2020}) often do not produce sufficient feedback at large halo radii to explain the $S_8$ tension and kSZ measurements \citep{Bigwood2024,McCarthy2024} (see, however, \citealt{Bigwood2025}).   This may be because not all the relevant feedback physics and processes are  included in the current generation of cosmological simulations.   In addition, it is also plausible that the dominant physics responsible for feedback in halo outskirts may be distinct from the physics that balances radiative cooling in the inner parts of halos probed by X-ray observations.

In this paper we consider feedback from the cumulative effect of relativistic CRs produced during the growth of massive BHs.   This feedback mechanism  (1)  is not included in most current cosmological simulations, particularly large-volume simulations used for cosmology (2)   preferentially acts in $\sim 10^{13-14} M_\odot$ halos (but not in more massive clusters), and (3) is not directly correlated with the current radiative cooling rate of hot gas in massive halos.   The presence of an energetically important population of relativistic CRs is an inevitable consequence of relativistic jets and strong winds produced by massive BHs.   The feedback produced by CRs is, however, quite distinct from, and in addition to, the mechanical feedback produced by jets  and winds as they shock and heat the ISM/CGM/ICM.   In particular, CR feedback remains dynamically important even well after the mechanical feedback generating the CRs has ceased.

\begin{figure}[ht]
\centering
\includegraphics[width=\columnwidth]{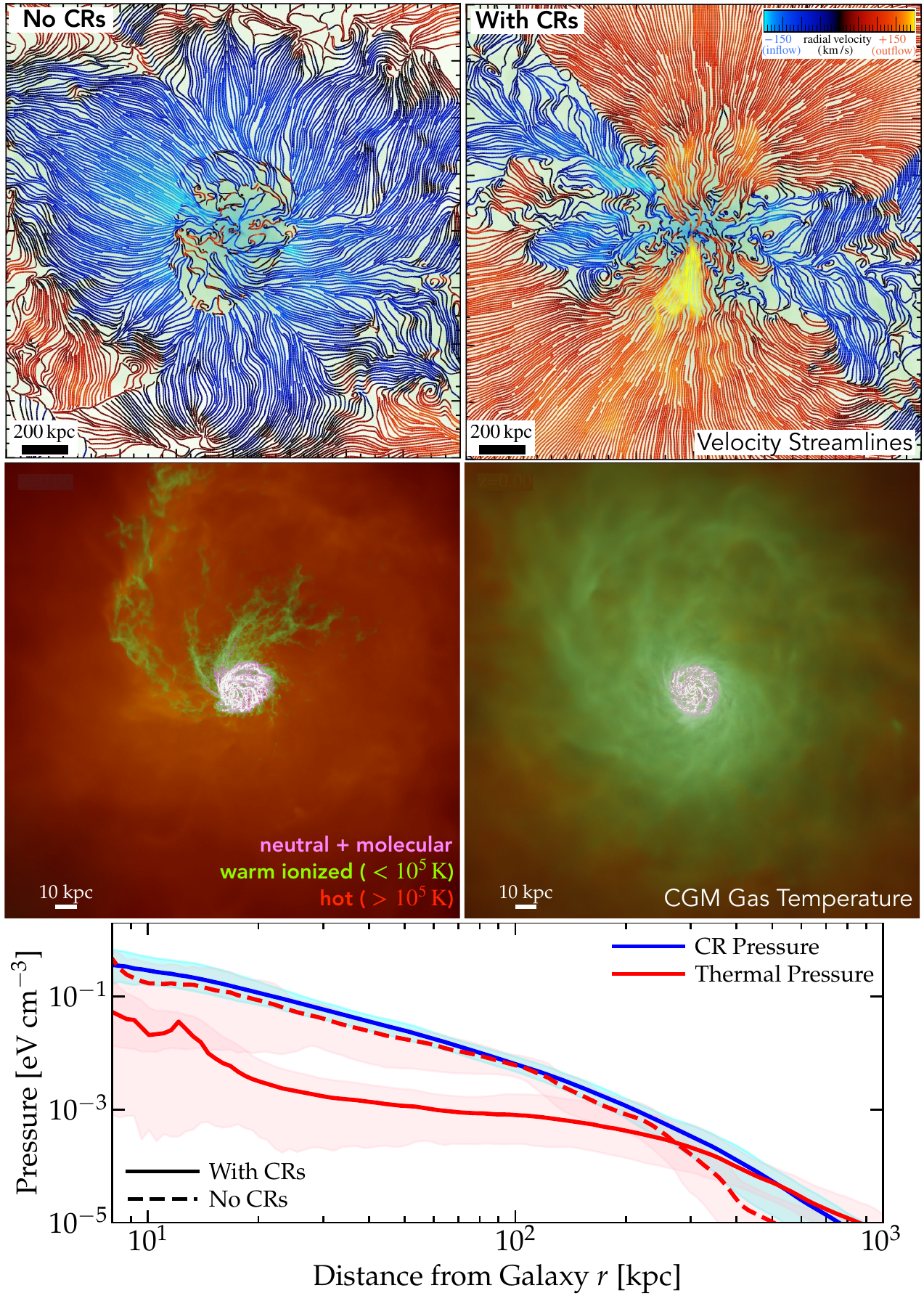}
\caption{
Properties of the circumgalactic medium in two zoom-in cosmological simulations of a $10^{12} M_\odot$ halo hosting a Milky-way mass galaxy (based on \citealt{Ji2020,Ji2021, Hopkins2021}). Neither includes black holes.  The left column in the images includes stellar feedback and magnetic fields while the right column (With CRs) also incudes CRs generated by core-collapse supernovae with a constant CR diffusion coefficient of $3 \times 10^{29} \dun$ and streaming at the Alfv\'en speed.  In the bottom panel the shaded range shows dispersion in pressure at a given radius while the lines show the spherical average.   The inclusion of CRs dramatically changes the dynamics and thermodynamics of the CGM, even well outside the virial radius:  the halo gas is predominantly photoionized and supported by CR pressure rather than thermal pressure and cosmological inflow of gas is replaced by CR-driven outflows of gas over a large fraction of the halo solid angle, with a significantly less prominent virial shock.   In this paper we propose that CRs produced by BHs can produce a similarly large impact on the halo gas in group-mass halos, although primarily at large radii rather than throughout the halo. \label{fig:CGMCR}}
\end{figure}

The goal of this paper is to motivate the potential importance of CR feedback in group-mass halos.   Given the uncertainties in CR transport (see \S \ref{sec:transport}) we cannot definitively establish that CR feedback is important, but order of magnitude considerations  suggest that it may be:  under a range of assumptions about CR physics, the pressure of relativistic CRs can be comparable to the thermal gas pressure in $\sim 10^{13-14} M_\odot$ halos.  A useful analogy  is that  simulations of the effects of CRs produced by core-collapse supernovae in $\sim 10^{12} M_\odot$ halos have already demonstrated that CRs can dramatically change the structure and observational properties of the circumgalactic medium, both inside and outside the virial radius (see Fig. \ref{fig:CGMCR}):  halos can become dominated by CR pressure rather than thermal pressure,  with cosmological inflow of gas replaced by CR-driven outflows of gas, and with a significantly less prominent virial shock \citep{Butsky2018,Ji2020,Ji2021, Hopkins2021,Butsky2022}.  The extended X-ray emission from halos with significant CR pressure is primarily  Inverse Compton upscattering of CMB photons by CR electrons, rather than  thermal emission \citep{Hopkins2025}.   It is not guaranteed that these changes to our standard picture of the CGM occur, however, because of uncertainties in CR transport: some alternative CR transport models show a significant, but somewhat less prominent, effect of supernovae-generated CRs on the structure of the circumgalactic medium \citep{Butsky2018,Hopkins2021c,Ramesh2024}.   
Because the total energy in CRs generated by BHs may well be  larger than that associated with supernovae (Fig. \ref{fig:CRtot} discussed below) it is plausible that BH-generated CRs can have a comparably important effect on the structure of hot gas in massive halos.   

The remainder of this paper is organized as follows.  We first motivate the energetic importance of CRs in massive halos, discuss their  spatial distribution, and quantify the conditions under which CR pressure is dynamically important in halo outskirts (\S \ref{sec:CRfeedback}).  The latter depends on the uncertain physics of CR transport (i.e., how CRs travel through the surrounding medium once they are produced), which we discuss in \S \ref{sec:transport}. In \S \ref{sec:NT}  we discuss the (not very good) prospects of directly detecting non-thermal emission from CRs in the outskirts of massive halos.   \S \ref{sec:disc} summarizes our results, discusses observational tests, and discusses the prospects for simulating the role of CR feedback in group mass halos.

\section{Cosmic Ray Feedback}
\label{sec:CRfeedback}
\subsection{Overall Energetics}
\label{sec:energetics}
\begin{figure}[ht]
\centering
\includegraphics[width=\columnwidth]{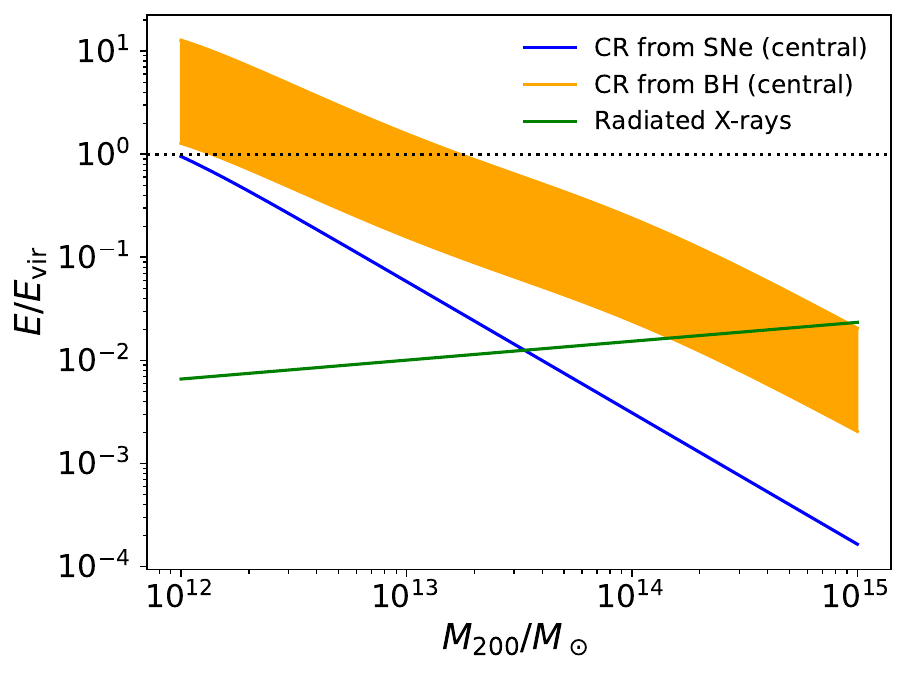}
\caption{
Estimates of various total energy components in massive halos relative to the thermal energy of a virialized halo with cosmic baryon fraction.  The shaded range for CRs from the central BH corresponds to CR energy injection fractions of $10^{-3}-10^{-2} M_{\rm BH} c^2$.  The radiated X-rays uses the $L_X-M_{200}$ correlation and assumes a similar X-ray luminosity over the last $5$ Gyrs.   In halos with masses $\gtrsim 10^{14} M_\odot$ the stellar (and thus BH) mass in non-centrals (i.e,. the rest of the group/cluster) is a factor of $\sim 2-4$ larger than in the central \citep{Kravtsov2018}, likely increasing the CR energetics by a similar factor.
\label{fig:CRtot}}
\end{figure}

To start, we motivate why CRs from BHs should be energetically important in massive halos, particularly those with masses $M_{\rm 200}\sim 10^{12-14} M_\odot$.  To estimate the typical central stellar and BH mass in a given halo of mass $M_{200} \gtrsim 10^{12} M_\odot$ (defined relative to critical) we use the $z \sim 0$ stellar mass halo mass relation from \citet{Moster2013}, {the bulge to total (B/T) stellar mass ratio from \citet{Fu2022}}, and the BH-stellar mass relation from \citet{Kormendy2013}.   {For massive halos with ${\rm B/T} \sim 1$, this gives $M_\star  \simeq 10^{11} M_\odot (M_{200}/10^{13} M_\odot)^{\alpha_\star}$ and $M_{BH}  \simeq 5 \times 10^{8} M_\odot (M_{200}/10^{13} M_\odot)^{\alpha_{BH}}$ with $\alpha_\star \simeq 0.4$ and $\alpha_{BH} \simeq 0.45$. The plot in Figure \ref{fig:CRtot} below includes the B/T factor which decreases the typical BH mass in $10^{12} M_\odot$ halos by a factor of 4 but is a negligible correction for $\gtrsim 10^{13.5} M_\odot$ halos.}
These estimates neglect the BHs and stars not in the central galaxy, i.e., those in the rest of the group/cluster. For $\sim 10^{14} M_\odot$ halos, we thus likely underestimate the net stellar and BH feedback energy by a factor of $\sim 2$ \citep{Kravtsov2018}.  This correction is smaller (larger) for lower (higher) halo masses.

Star formation produces CRs primarily via core-collapse supernovae\footnote{Integrated over a Hubble time, Ia supernovae contribute to the total CR budget at the $\sim 10\%$ level, while fast stellar winds, colliding wind binaries, pulsars, and X-ray binaries are even smaller sources.} which produce $\sim 10^{50} \epsilon_{\star,-1}$ ergs of CR energy per supernovae, where $\epsilon_\star = 0.1 \epsilon_{\star,-1}$ is the fraction of a supernova's energy that goes into CRs.   Given a typical stellar initial mass function there is roughly 1 supernova per 100 $M_\odot$ of stars formed, implying that the total CR energy from supernovae over the history of star formation in a massive halo is 
\begin{equation}
E_{\rm CR,SNe} \sim 10^{59} \, {\rm erg} \, \epsilon_{\star,-1} \left(\frac{M_{200}}{10^{13} M_\odot}\right)^{0.4}
\label{eq:EtotCRSNe}
\end{equation}
The CR energetics from BH growth is somewhat less  well-constrained. CRs are produced both by AGN-driven winds shocking with ambient gas and via relativistic jets.   We assume that a fraction $\epsilon_{BH} = 10^{-3} \epsilon_{BH,-3}$ of the BH's rest mass energy goes into CRs and we take $\epsilon_{BH} \sim 10^{-2}-10^{-3}$.   The motivation for this choice is as follows.  Theoretically, in zoom-in simulations of massive halos, \citet{Wellons2023} found that $\epsilon_{BH} \sim 10^{-3}$ in CRs  was capable of quenching massive galaxies in halos $\sim 10^{13} M_\odot$. Observationally, $\sim 10\%$ of quasars are radio-loud and the jet power in radio-loud systems is believed to be (largely on theoretical grounds) of order $\dot M c^2$ (e.g., \citealt{Tchekhovskoy2011}).  This would imply a total jet energy over the course of BH growth of $\sim 10^{-2} M_{BH} c^2$ or even more.  Most of the jet energy is dissipated via a strong shock at the head of the jet as it interacts with the ambient medium and/or via magnetic instabilities in the jet.   The jet also deposits some of its energy into shocks in the ambient medium driven by the over-pressured cocoon of shocked-jet material \citep{Begelman1989}.  
These processes will convert most of the jet energy into relativistic particles; $\epsilon_{BH} \sim 10^{-3}-10^{-2}$ are thus plausible.   Quasar-driven winds with speeds $\sim 0.03-0.3$ c (e.g., broad-absorption line quasar winds, ultra-fast outflows) are probably a subdominant but non-negligible source of CRs as well.   Our rough estimate of the total energy in CRs supplied by the growth of the central BH in a given halo is thus
\begin{equation}
E_{\rm CR,BH} \sim 10^{60} \, {\rm erg} \, \epsilon_{BH,-3} \left(\frac{M_{200}}{10^{13} M_\odot}\right)^{0.45}
\label{eq:EtotCRBH}
\end{equation}
The CR energy from BHs is thus likely to be at least an order of magnitude larger than the energy in CRs from supernovae (in the massive, early-type galaxies of interest).  

Figure \ref{fig:CRtot} compares the CR energetics estimated above to the total thermal energy of a virialized halo $E_{vir}$ with cosmic baryon fraction and to the radiated energy in X-rays over the last 5 Gyrs (using $L_X(M_{200}$) from \citealt{Anderson2015}).   Two things are striking:    the first is that the CR energy supplied by BH growth is of order or exceeds the {\em total} virialized thermal energy of the halo for $\sim 10^{12-13.5} M_\odot$ halos.  It is difficult to avoid the conclusion that CR feedback will be significant in these systems. This is particularly true given the numerical evidence that CRs from core-collapse supernovae alone can change the halo gas dramatically in $10^{12} M_\odot$ halos when $E_{CR,\star} \sim E_{vir}$ (Fig. \ref{fig:CGMCR}).  Given the similar energetics of CRs produced by BHs in $10^{13-14} M_\odot$ halos (Fig. \ref{fig:CRtot}), it is reasonable to expect that CRs from BHs will have a large impact on the halo gas.  

The second striking result in Figure \ref{fig:CRtot} is that the CR energy supplied during the primary epoch of BH growth greatly exceeds the total radiated thermal energy at low redshift in halos of $\lesssim 10^{14} M_\odot$.   Models of AGN feedback in massive halos typically focus on balancing the current radiative cooling in the halo and thus quenching star formation in massive galaxies.   While this is of course empirically well motivated, Figure \ref{fig:CRtot} strongly suggests that the cumulative impact of CRs produced at higher redshift during the era of massive black hole growth is likely to be as or more important than the $z \sim 0$ feedback (particularly in group-mass halos).

\subsection{Spatial Distribution of Cosmic-Rays}
\label{sec:spatial}

The impact of CRs in massive halos will of course depend on their spatial distribution, which in turn depends on the uncertain physics of CR transport.  Here we illustrate the possible impact of CRs using order of magnitude estimates.   We model the CR spatial distribution assuming they undergo isotropic spatial diffusion with a constant diffusion coefficient $\kappa$.  In \S \ref{sec:transport} and Appendix \ref{sec:app}, we return to the uncertainties in CR transport, the role of streaming vs. diffusion, and the likely dependence of the diffusion coefficient on halo mass. We choose to present our results in this section in terms of a CR diffusion coefficient in part because there are simple known analytic solutions for diffusive transport and because diffusive transport is likely to be more familiar to the reader not versed in CR transport theory than streaming transport.  

The results that follow also neglect significant loss of CR energy from production in jets and/or quasar winds to the outer parts of halos.  This approximation is most readily justified if the bulk of the CRs in massive halos are protons (as is the case in the Milky Way) because CR proton loss times are long.  {The dominant collisional/radiative loss process for $\gtrsim$ GeV protons is pion decay, for which the energy loss timescale is $\simeq 10^{10} (n/0.01 \, {\rm cm^{-3}})$ yrs for an ambient gas density $n$ scaled to a typical value in groups and clusters at $\sim 0.1 R_{200}$; this is much longer than the transport timescale in the models considered here.}  However, the composition of relativistic jets is still uncertain and they may instead be primarily electron/positron pairs (e.g., \citealt{blandford:2019.jet.review}), which have much higher radiative and Coulomb losses. We return to this uncertainty in \S \ref{sec:disc}.  One important distinction between diffusive and streaming transport is that in the latter case the energy in CRs is not conserved because CRs continuously heat the thermal plasma (the heating is mediated by damping of the CR-streaming-generated waves; e.g., \citealt{Wentzel1971,Zweibel2017}).   We show in Appendix \ref{sec:app}, however, that the net loss of energy via this mechanism is modest in our application and so reasonable to neglect for the purposes of these initial estimates.

Given these simplifying assumptions, the CR distribution at late times (i.e., lowish redshift) and at large radii in halos can be reasonably approximated by simply assuming that the majority of the CRs are produced in the center of the halo at early times.   The solution of the diffusion equation for a delta function source in both time and space thus yields
\begin{equation}
p_{CR}(r,t) \simeq \frac{E_{CR,BH}}{3 (4 \pi \kappa t)^{3/2}} \exp(-r^2/4 \kappa t)
\label{eq:PCRdeltafcn}
\end{equation}
This ignores the cosmological assembly of the halo but the uncertainties in CR transport are larger than those introduced by not properly modeling structure formation.

\begin{figure}[ht]
\centering
\vspace{0.5cm}
\includegraphics[width=\linewidth]{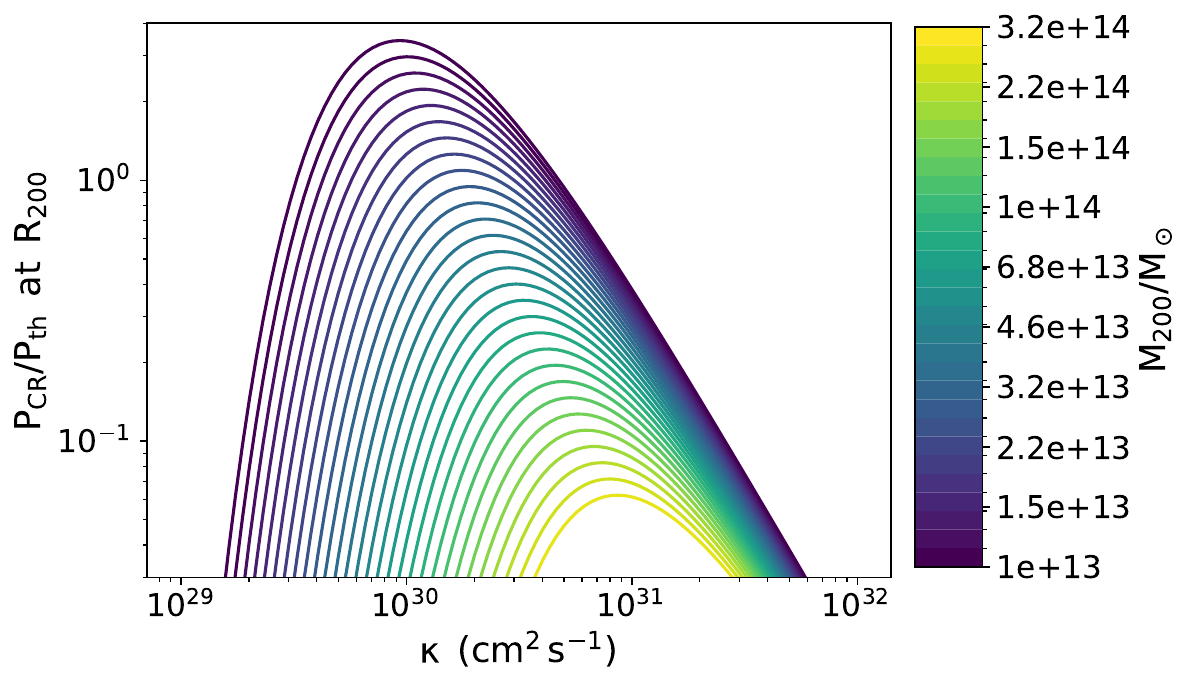}
\caption{Estimate of the dynamical importance of CRs near the virial radius as a function of CR diffusion coefficient and halo mass.  We show the ratio of CR to thermal pressure at $R_{200}$ assuming the self-similar thermal pressure from \citet{Arnaud2010} and  assuming CRs were injected $\sim 10$ Gyr ago during the primary epoch of massive black hole growth.  The curves assume $\epsilon_{CR,BH} = 3 \times 10^{-3}$ but the CR pressure can be scaled linearly for other values of $\epsilon_{CR,BH}$ (eq. \ref{eq:EtotCRBH}).   For a range of CR diffusion coefficients and halo masses CR pressure is dynamically important near the virial radius and is likely to drive baryons well outside the halo.  Note that $\kappa$ here is the effective $\kappa$ in the CGM/ICM at large radii $\sim R_{200}$, which is expected to be much larger than the Solar-neighborhood ISM value, and itself should depend on local plasma conditions and thus on halo mass (see \S \ref{sec:transport} and eq. \ref{eq:kappastr}).
\label{fig:CRR200}}
\end{figure}

To compare equation \ref{eq:PCRdeltafcn} to the thermal pressure of the halo gas, we assume that the halo baryons have the universal pressure profile of \citet{Arnaud2010}, which is derived empirically as function of halo mass and radius (primarily in massive systems like galaxy clusters).  Figure \ref{fig:CRR200} shows the resulting ratio of the CR pressure to the thermal gas pressure at $r = R_{200}$ as a function of halo mass and CR diffusion coefficient (we assume $t \sim 10$ Gyr, consistent with most BH growth at $z \sim 1-2$).   Figure \ref{fig:CRR200} assumes $\epsilon_{CR,BH} = 3 \times 10^{-3}$ but the resulting $p_{CR}/p_{gas}$ can be scaled linearly for any other value of $\epsilon_{CR,BH}$.    Note that the diffusion coefficient $\kappa$ in Figure \ref{fig:CRR200} is the effective  $\kappa$ in the CGM/ICM at $\sim R_{200}$, which is expected to be much larger than the Milky Way ISM value (see \S \ref{sec:transport}), and itself will depend on local plasma conditions and thus on halo mass.

Figure \ref{fig:CRR200} shows that for a wide range of CR diffusion coefficients $\sim 3 \times 10^{29} \, {\rm cm^2 s^{-1}} \lesssim \kappa \lesssim 10^{31} \, {\rm cm^2 s^{-1}}$ there are halo masses in which $p_{CR} \sim p_{gas}$ at large halo radii; taking smaller (larger) $\epsilon_{CR,BH}$ would, of course, decrease (increase) this range of diffusion coefficients.    Figure \ref{fig:CRR200} also reinforces that CRs are likely to be dynamically important primarily for halo masses $\lesssim 10^{14} M_\odot$ simply because for the most massive halos the CRs are an energetically less important constituent (Fig. \ref{fig:CRtot}).  The scalings with diffusion coefficient and halo mass in Figure \ref{fig:CRR200} can be understood as follows.  For low diffusion coefficients in Figure \ref{fig:CRR200} the CRs cannot diffuse to the virial radius in a Hubble time and so $p_{CR}/p_{gas} \ll 1$ at $r = R_{200}$.   For sufficiently large CR diffusion coefficients by contrast, $p_{CR} \propto E_{CR,BH} \kappa^{-3/2} \propto M_{200}^{0.45} \epsilon_{CR,BH} \kappa^{-3/2}$.   The thermal pressure at the virial radius in \citet{Arnaud2010}'s model roughly follows the self-similar scaling $p_{gas} \propto M_{200}^{2/3}$ so that $p_{CR}/p_{gas} \propto \epsilon_{CR,BH} M_{200}^{-0.2} \kappa^{-3/2}$.   

\begin{figure}[ht]
\centering
\includegraphics[width=\linewidth]{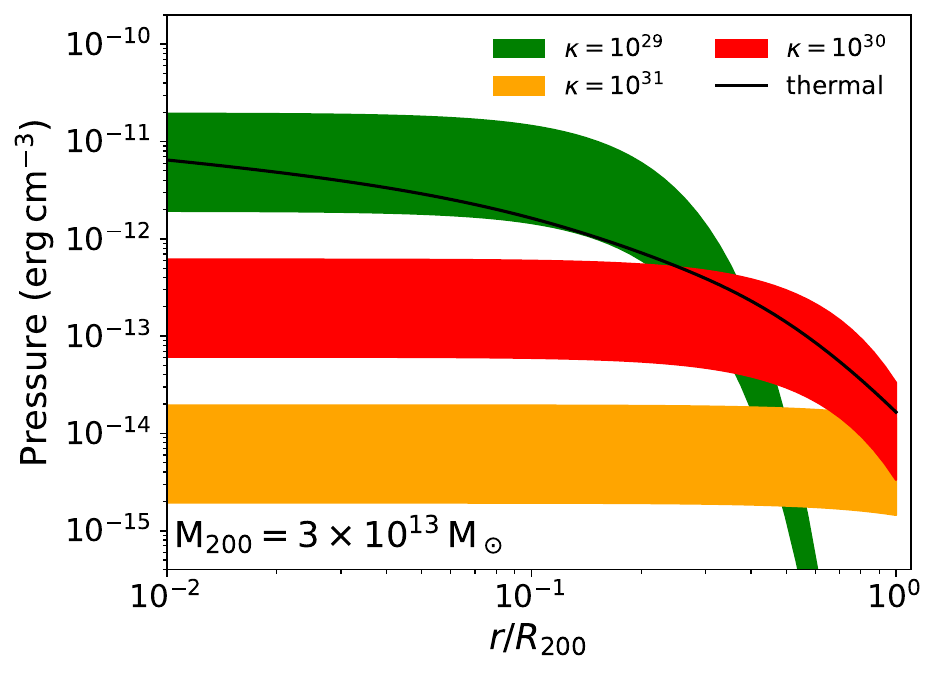}
\includegraphics[width=\linewidth]{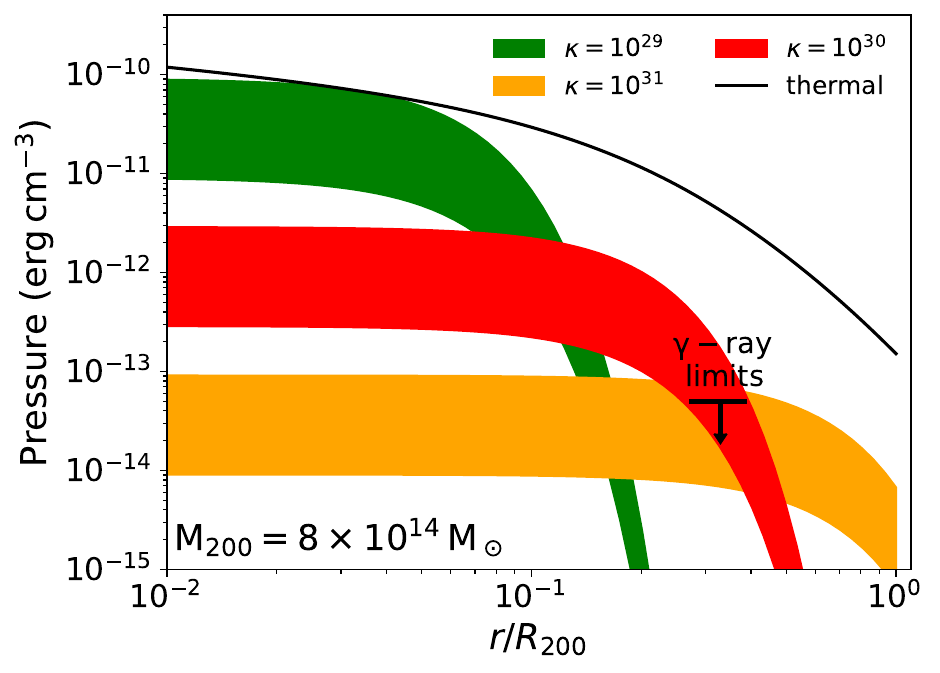}
\caption{Thermal pressure profile in a group (top) and cluster (bottom) mass halo compared to CR pressure profiles for CRs injected $\sim 10$ Gyr ago which have since diffused to larger radii.  The shaded range covers CR energy injection fractions of $10^{-3}-10^{-2}$.   The upper limits in the cluster panel are from the lack of pion decay gamma-rays in Fermi data of a stacked sample of 50 galaxy clusters \citep{Huber2013,Ackerman2014}; this favors larger diffusion coefficients, at least in high mass systems.
\label{fig:CRradial}}
\end{figure}

Figure \ref{fig:CRradial} shows radial profiles of the CR pressure (eq. \ref{eq:PCRdeltafcn}) and thermal gas pressure \citep{Arnaud2010} for $3 \times 10^{13} M_\odot$ (top) and $8 \times 10^{14} M_\odot$ (bottom) halos for several CR diffusion coefficients.  The shaded range of $p_{CR}$ corresponds to $\epsilon_{CR,BH} = 10^{-3}-10^{-2}$.  For group-mass systems with diffusion coefficients $\gtrsim 10^{30-31} {\rm cm^2 s^{-1}}$ the CR pressure is modest compared to the thermal gas pressure throughout the inner halo.   However, at large radii, the gas pressure drops significantly and CRs can become dynamically important.   This is more likely to occur in group mass halos than cluster mass halos given the larger overall ratio of CR to thermal energy in lower mass halos (Fig \ref{fig:CRtot}).    

Figure \ref{fig:CRradial} shows that if $\kappa \sim 10^{29} \dun$ then the CR pressure could dominate the gas pressure even in the inner parts of massive halos.   For a few clusters at radii of $\sim 10$ kpc this is constrained by comparisons of stellar and X-ray inferred gravitational potentials \citep{Churazov2008}.  Gravitational lensing and X-ray inferred total masses of clusters (e.g.\ $M_{500}$ or $M_{200}$) also generally agree to $\sim 10s \%$ (e.g., \citealt{Zhang2008,Simet2017}), appearing to rule out $p_{CR} \gtrsim p_{th}$.   The most direct constraint on CRs in massive halos comes from $\gamma$-ray observations.   In particular, Fermi has searched extensively for gamma-ray emission from pion decay in massive clusters, as a probe of the CR proton energy content in the intracluster medium, {with at most one unambiguous detection of extended diffuse gamma-ray emission (e.g., \citealt{Huber2013,Ackerman2014}); the likely detection is Coma \citep{Adam2021,Baghamanyan2022}.}   The gamma-ray luminosity $\propto \int n_{gas}(r) p_{CR}(r) r^2 dr$ and so provides a constraint on $p_{CR}$ given that the gas density in these systems is measured through X-ray observations.  The gamma-ray upper limits correspond to upper limits on the mean CR pressure fraction (averaged out to the virial radius) in massive clusters ($M_{200} \sim 8 \times 10^{14} M_\odot$ is the average cluster mass in the stacked samples of \citealt{Huber2013,Ackerman2014}).   The exact upper limit on $p_{CR}/p_{gas}$ depends on a number of assumptions, including the assumed radial CR pressure profile, the gas densities at large radii, and the energy spectrum of CR protons.  For the specific theoretical models in \citet{Huber2013,Ackerman2014}, the authors inferred $p_{CR}/p_{gas} \lesssim 0.02-0.04$ at large radii in massive clusters {(the Coma detection, if indeed due to pion decay, corresponds to $p_{CR}/p_{gas} \sim 0.02$).}   We show this upper limit in Figure \ref{fig:CRradial} (bottom panel) at radii $\sim 0.3 R_{200}$ that should typically dominate the (unresolved) gamma-ray emission given typical cluster gas density profiles.   Observational constraints in massive clusters thus imply that cosmic-rays have not had a large dynamical effect on the baryon distribution at the virial radius ($R_{200}$) in massive halos (although the observations do not constrain strong CR effects earlier during halo formation/assembly, and only weakly constrain strong CR effects in cool core interiors).   However, Figs.~\ref{fig:CRR200} and \ref{fig:CRradial} show that this is not true in less massive systems.   CR diffusion coefficients of $\sim 10^{30-31} \dun$ are capable of producing CR pressures of order the thermal gas pressure at large radii in groups while at the same time being consistent with gamma-ray upper limits in galaxy clusters.   There is also no reason to expect that the CR diffusion coefficient is the same in lower mass and massive systems (and certainly not the same as in the local Milky Way ISM).  On the contrary, CR transport depends on the local plasma properties (e.g., \citealt{Ruszkowski2023}) and so may well change systematically with halo mass,  as we show in \S \ref{sec:transport}.

Fermi gamma-ray upper limits on the CR content of massive halos are the most constraining for the most massive systems because the gamma-ray luminosity scales with the total baryonic mass at fixed $p_{CR}$.  There are, however, several fortuitously nearby $\sim 5 \times 10^{13} M_\odot$ groups in the Fermi sample of \citep{Huber2013,Ackerman2014}, e.g., NGC 5044 and NGC 1550.   The upper limit on the average CR pressure at $\sim R_{200}$ (assuming a universal gas density profile) is roughly $p_{CR} \lesssim 10^{-13} {\rm erg \, cm^{-3}}$ in these systems.  This again favors larger diffusion coefficients given the models in Figure \ref{fig:CRradial} but leaves a large parameter space of models in which CR pressure can be dynamically important near the virial radius or beyond. An interesting question is how strongly this current upper limit on the CR content of a few groups constrains the integrated effect of CR feedback over the history of the group; assessing this will require cosmological simulations including CRs (see \S \ref{sec:disc}).  

{One uncertainty in assessing the effect of CRs on baryons in massive halos is that at $z \sim 1-2$, when the CRs were originally produced, their pressure in the inner halo was likely much larger than it is at lower redshift.   This follows simply from energy conservation, as the CR pressure must be larger at earlier times when the CRs were contained in a smaller volume.  To see this in the context of our analytic models, note that the solution for CR pressure in equation \ref{eq:PCRdeltafcn} depends on time only through the combination $\kappa t$.  Thus the factor of 100 variation in CR diffusion coefficient in Figure \ref{fig:CRradial} can equivalently be viewed as a factor of 100 variation in time since the CRs were produced. That is, the green curves equally well describe the CR pressure $10^{10}$ yrs after the CRs were produced with $\kappa = 10^{29} \, \dun$ and the CR pressure $10^8$ yrs after the CRs were produced with $\kappa = 10^{31} \, \dun$.   Naively, models with large pressure at lower redshift in the outer halo thus inevitably had significant CR pressure in the inner halo at earlier times when the BH was active.   Cosmological simulations will be required to assess whether models with strong CR feedback in the outer halo can indeed produce groups consistent with X-ray observations of the inner halo.  We  return to this point in more detail in \S \ref{sec:disc}.}

It is useful to assess whether the dominant source of CR pressure at large radii in groups is likely to be from CRs produced early during the phase when the BH gained most of its mass (eq. \ref{eq:PCRdeltafcn}) or later when ongoing BH feedback at low redshift helps keep the central galaxy quiescent (this is effectively the question of whether early `quasar' mode or later `maintenance' mode feedback is the most important source of CRs considered in this paper).   To estimate the CR pressure profile produced by  low redshift BH accretion and feedback, we consider a simple model in which BH accretion at lower redshift supplies a constant CR power $\dot E_{CR}$ over a time $t_{on}$, in which case
\begin{equation}
p_{CR}(r,t) \simeq \frac{\dot E_{CR}}{12 \pi \kappa r} \left(1 -{\rm Erf}\left[\frac{r}{\sqrt{4 \kappa t_{on}}}\right]\right)
\label{eq:PCREdot}
\end{equation}
If we assume that $\dot E_{CR} = f L_X$ with $f \sim 1$ motivated by models of self-regulated jet feedback in massive halos (e.g., \citealt{Fabian2012, Li2014}), we find that the ratio of the CR pressure at $R_{200}$ from `early' feedback (eq. \ref{eq:PCRdeltafcn}) to the CR pressure at $R_{200}$ from `maintenance' feedback (eq. \ref{eq:PCREdot}) is
\begin{equation}
\frac{\rm p_{CR, early}}{\rm p_{CR,maintenance}} \sim \frac{30 \, \epsilon_{BH,-3}}{f} \left(\frac{M_{200}}{10^{13} M_\odot}\right)^{-1} \left(\frac{\kappa}{10^{30} \, \dun}\right)^{-1/2}
\label{eq:pcratio}
\end{equation}
where we have used the observed $L_X \propto M_{200}^{1.85}$ relation from \citet{Anderson2015} and have assumed that $\kappa$ is large enough that  finite diffusion time effects do not strongly suppress the CR pressure at large radii (and thus we can neglect the exponential and error function suppression terms in eq. \ref{eq:PCRdeltafcn} and \ref{eq:PCREdot}, respectively).  Equation \ref{eq:pcratio} shows that in group mass halos the CR pressure at large radii is indeed likely to be dominated by the cumulative impact of CRs produced at early times, not the CRs produced by current BH activity.

\vspace{1cm}

\section{CR Transport}
\label{sec:transport}

The physical processes regulating CR transport are still not fully understood (see, e.g., \citealt{Ruszkowski2023} for a review).   Theoretically, CRs scatter off of magnetic-field fluctuations with a spatial scale comparable to the proton Larmor radius; this is $r_L \sim 3 \times 10^{12} B_{\mu G}^{-1}$ cm for a GeV proton in a 1 $B_{\mu G} \ {\rm \mu G}$ magnetic field, i.e., $\sim 10^{-12} R_{200}$ in groups!  These small-scale fluctuations  can either be the small-scale tail of a turbulent cascade (e.g., \citealt{Yan2002}), intermittent small-scale reversals in magnetic field direction (e.g., \citealt{Kempski2023,Lemoine2023,Butsky2024}), or magnetic-field fluctuations generated by the CRs themselves (e.g., \citealt{Kulsrud1969}).  If CRs are not efficiently scattered by ambient turbulence, any net drift of the cosmic-rays exceeding the Alfv\'en speed $v_A$ will excite the gyro-resonant streaming instability \citep{Lerche1967}; if these short-wavelength Alfv\'en waves are only weakly damped (see below), they can grow to the point that they can scatter the cosmic-rays and limit the resulting CR streaming speed to be similar to the Alfv\'en speed (e.g., \citealt{Bai2019}).  In general, there are diffusive corrections to streaming at the Alfv\'en speed (e.g. \citealt{Skilling1971}) so that both contributions to the CR energy flux can be important.\footnote{The distinction between streaming and diffusive transport mathematically corresponds to a CR energy flux $F_c$ given by (see Appendix \ref{sec:app} for details)
\begin{equation}
{\bf F}_c=4 p_{CR} {\bf v_{\rm s}}  \ \ \ {\rm or} \ \ \ 
{\bf F_c} = -\kappa \,{\bf n}\left({\bf n}\cdot\nabla E_{CR}\right), \nonumber
\end{equation}
where ${\bf v_s} = -{\bf v_A}\, ({\bf n} \cdot \nabla p_{CR}) /|\nabla p_{CR}|$ is the streaming speed down the CR pressure gradient, $p_{CR} \simeq E_{CR}/3$, ${\bf v_A}={\bf B}/(4\pi\rho)^{1/2}$ is the Alfv\'en speed, and ${\bf n}={\bf v_{\rm A}}/|\bf v_{\rm A}|$.    These expressions are valid only on scales larger than the mean-free-path of the $\sim$ GeV energy CRs that dominate the total energy of the CR population, so that a fluid description, rather than a fully kinetic one, is applicable.}

Our goal in this section is to provide order of magnitude estimates to assess whether CR diffusion coefficients of order those needed to produce $p_{CR} \sim p_{gas}$ in group-mass halos are plausible or not.   The most well-understood CR transport mechanism is arguably that produced by waves generated by the streaming instability.    If CRs collectively stream down their pressure gradient at a speed $v_D$ the growth rate of resonant Alfv\'en waves with wavelengths $\sim r_L$ due to the streaming instability is (e.g., \citealt{Zweibel2017})
\begin{equation}
\Gamma_{str} \sim 3 \Omega_p \frac{p_{CR}}{p_{gas}} \frac{kT}{m_p c^2} \left(\frac{v_D}{v_A} - 1\right)
\label{eq:growth}
\end{equation}
where here and in what follows we will focus on the CRs that dominate the CR pressure, which are those with energies of $\sim$ GeV.   In equation \ref{eq:growth}, $\Omega_p$, is the non-relativistic proton cyclotron frequency and $v_A$ is the Alfv\'en speed.  The importance of the streaming instability depends on the damping rate of the excited waves $\Gamma_{damp}$ relative to the driving rate $\Gamma_{str}$.   \citet{Wiener2013} and \citet{Wiener2018} discuss the dominant damping in the diffuse hot plasmas relevant to groups and clusters and argue that it is collisionless damping of the excited waves \citep{Wiener2018}.   The resulting damping rate is 
\begin{equation}
\Gamma_{damp} \sim \beta^{1/2} \frac{\sqrt{\pi}}{4} \left(\frac{\dot e_{turb}}{v_A r_L}\right)^{1/2} \sim \frac{\sqrt{2 \pi}}{4} \frac{c_s}{\sqrt{L_{MHD} r_L}}
\label{eq:damp}
\end{equation}
where $\dot e_{turb} \sim v_A^3/L_{MHD}$ is the turbulent energy cascade rate in the ambient medium, $L_{MHD}$ is the outer scale of such turbulence,\footnote{The ambient turbulence is important because it generates a perpendicular variation associated with the waves excited by the streaming instability, which makes the waves susceptible to collisionless damping \citep{Wiener2018}.} $c_s$ is the gas sound speed, and $\beta = 2 c_s^2/v_A^2$ is the ratio of gas to magnetic pressure.   Balancing driving and damping we find that the equilibrium drift speed reached by the CRs in the presence of the streaming instability is
\begin{equation}
\frac{v_D}{v_A} - 1 \sim 0.2 \, \frac{p_{gas}}{p_{CR}} \frac{c}{c_s} \sqrt{\frac{r_L}{L_{MHD}}} 
\label{eq:drift1}
\end{equation}
To provide an order of magnitude estimate we assume that $c_s$ is of order the virial velocity in a halo of mass $M_{200}$ and that $L_{MHD} \sim 0.1 \ell_{0.1} R_{200}$, in which case
\begin{equation}
\frac{v_D}{v_A} - 1 \sim 10^{-3} \, \frac{p_{gas}}{p_{CR}} \left(\frac{M_{200}}{10^{13} M_\odot}\right)^{-1/2} \ell_{0.1}^{-1/2} B_{\mu G}^{-1/2}
\label{eq:drift2}
\end{equation}

Equation \ref{eq:drift2} motivates that unless $p_{gas} \gg p_{CR}$ the streaming instability may be able to regulate $v_D \sim v_A$.   Since we are interested in assessing whether the regime with $p_{CR} \sim p_{gas}$ is realizable at some radii, we assume that this is true and assess its implications.   If the CRs stream at a drift speed of $v_A$ this is analogous to diffusing with $\kappa \sim r v_A$ (such the streaming time is of order the diffusion time $r/v_A \sim r^2/\kappa$, or equivalently, that the streaming energy flux is of order the diffusive energy flux [see footnote 1 and Appendix \ref{sec:app}]).   Note that this identification applies most directly to the field-aligned transport of CRs.  The effective isotropic diffusivity used in our analytic estimates in \S \ref{sec:CRfeedback} is likely a factor of few less than the field-aligned transport coefficient because of the tangled/chaotic magnetic field structure \citep{Narayan2001}.   Neglecting this factor, i.e., considering the field-aligned transport coefficient, our order of magnitude estimate of the effective diffusion coefficient associated with CRs streaming at the Alfv\'en speed is thus
\begin{equation}
\kappa_{eff,str} \sim 6 \times 10^{31} \, \dun \, \beta^{-1/2} \left( \frac{r}{R_{200}}\right) \, \left(\frac{M_{200}}{10^{13} M_\odot}\right)^{2/3}.
\label{eq:kappastr}
\end{equation}
{The $M^{2/3}$ scaling here follows from self-similarity of structure formation:  for streaming, $\kappa_{eff,str} \sim r v_A \sim r c_s \beta^{-1/2}$.  Typical cluster radii scale as $r \propto R_{200} \propto M_{200}^{1/3}$ and typical sound speeds scale as $c_s \propto \sqrt{M_{200}/R_{200}} \propto M_{200}^{1/3}$.}  Equation \ref{eq:kappastr} is, to order of magnitude, similar to the diffusion coefficients that would produce $p_{CR} \sim p_{gas}$ in the outskirts of group mass halos (Figs \ref{fig:CRR200} and \ref{fig:CRradial}).  E.g., if $\beta \sim 100$, $\kappa \sim 2 \times 10^{30} \, \dun \, (M_{200}/10^{13} M_\odot)^{2/3}$ at $r = R_{200}/3$.   This would lead to $p_{CR} \sim p_{gas}$ at large radii in groups while satisfying $p_{CR} \ll p_{gas}$ in cluster mass halos (Figs \ref{fig:CRR200} \& \ref{fig:CRradial}).    The order of magnitude plausibility of the needed diffusion coefficients motivates more detailed calculations.   It is also interesting to note that applied to the inner CGM ($r \sim 0.05 R_{200}$) of MW-mass (10$^{12} M_\odot$) halos, equation \ref{eq:kappastr} corresponds to $\kappa \sim 6 \times 10^{29} \beta^{-1/2} \, \dun$, which is of order the value favored by phenomenological models of CR propagation in the MW calibrated to observational data (e.g., \citealt{Trotta2011}).  The dynamical impact of CRs may  thus extend well beyond the virial radius (as it does in the simulations in Fig. \ref{fig:CGMCR}).

Another way to think about CR transport via streaming in the context of halos is that the sound crossing time of a virialized halo is $\sim 0.1 \, t_{\rm Hubble}$.  The Alfv\'en crossing time is thus $\sim f_s^{-1} 0.1 \beta^{1/2} t_{\rm Hubble}$ where the effective radial speed is taken to be $f_s v_A$.  Depending on the exact values of $f_s$ and $\beta$ (which are uncertain), CRs streaming at $v_A$ can thus just reach a few times the virial radius  in a Hubble time.  The highest diffusion coefficient model in the upper panel of Figure \ref{fig:CRradial} thus requires $\beta^{1/2} f_s^{-1} \lesssim 10$ so that transport is in fact diffusive not purely streaming.   In general there are corrections to streaming at $v_A$ due to the damping of the waves scattering the CRs \citep{Skilling1971,Wiener2013}; these corrections always increase the effective transport speed of the CRs (so $f_s \gtrsim 1$ is in principle possible).  Even in the paradigm of scattering by waves generated by the streaming instability, transport can thus be more rapid than pure streaming at $v_A$.   Note also that if the CR pressure builds up to become dynamically important and drive an outflow then the dominant transport becomes advection with the outflowing gas which can be faster than the Alfv\'en speed (as in the well-developed theory of CR-driven galactic winds, where the flow transitions from sub to super-Alfv\'enic as a function of radius; e.g., \citealt{Ipavich1975,Mao2018,Quataert2022}).

In the context of streaming an alternative estimate of the dynamical importance of CR pressure near the virial radius  (analogous to Figure \ref{fig:CRR200}) is that we can roughly estimate the CR pressure at large radii as $p_c \sim E_{\rm CR,BH}/(4 \pi (v_s t)^3)$ for $r < v_s t$.   
Using $v_s = f_s v_A$ and estimating the sound speed as the virial velocity, this gives 
\begin{equation}
\frac{p_{\rm CR}}{p_{\rm th}}|_{r \sim R_{200}} \sim 4 \, \epsilon_{BH,-3} \left(\frac{M_{200}}{10^{13} M_\odot}\right)^{-7/6} \frac{(\beta/100)^{3/2}}{f_s^3 (t/10^{10} \, {\rm Gyr})^3}
\label{eq:pratstr}
\end{equation}
where we have assumed that $v_s t > R_{200}$ and have used the thermal pressure profile from \citet{Arnaud2010}, which satisfies the self-similar scaling $p_{th} \propto M^{2/3}_{200}$ at large radii.   Equation \ref{eq:pratstr} again suggests a dynamically important CR pressure at large radii in group mass halos, though the results are unfortunately sensitive to the uncertain microphysical parameters related to the streaming speed and magnetic field structure ($\beta$ and $f_s$).   The relatively strong halo mass dependence in equation \ref{eq:pratstr} also implies that CR pressure is unlikely to be important in massive galaxy clusters, consistent with observations.

The preceding estimates assume that ambient turbulent fluctuations do not efficiently scatter CRs, so that the primary source of pitch-angle scattering is the Alfv\'en waves generated by the CRs themselves via the streaming instability.   If there is also scattering by external turbulent fluctuations $\kappa_{ext}$ then the dominant source of scattering that produces a diffusion coefficient is the mechanism with the smaller diffusion coefficient:  the reason is that if $\kappa_{ext} \lesssim r v_A$, the CR distribution function will be sufficiently isotropic that it will be stable to the streaming instability so that the external turbulent source of scattering will set the CR transport; on the other hand, if $\kappa_{ext} \gtrsim r v_A$, the CR distribution function will be unstable to the streaming instability and so streaming will regulate the CR transport.   Unfortunately the scattering of CRs by external turbulence is less well-understood than scattering by the streaming instability:   scattering by Alfv\'enic and slow mode turbulence is ineffective \citep{Chandran2000} so theoretical models have focused on scattering by fast waves (e.g., \citealt{Yan2002}).   However, fast waves are strongly damped by wave steepening and other processes and so are unlikely to be efficient sources of CR scattering \citep{Kempski2022}.   Scattering by turbulence may instead be due to intermittent small-scale magnetic field reversals whose statistical properties are not well-understood \citep{Kempski2023,Lemoine2023,Butsky2024}, making reliable estimates of $\kappa_{ext}$ difficult at this time.   It is also possible that in the dilute plasmas present in massive halos, turbulent fluctuations associated with kinetic instabilities dominate CR scattering rather than the fluid turbulent fluctuations focused on in most of the CR transport literature \citep{Reichherzer2025}.

Although we have estimated diffusion coefficients in massive halos here in terms of global halo quantities such as distance from the halo center and halo mass (eq. \ref{eq:kappastr}) it is important to stress that these are really proxies for the transport depending on the local conditions in the plasma:  its ionization state, the statistical properties of turbulence, the magnetic field strength, the number of CRs, etc. (eq. \ref{eq:drift1}).  It is well known from cosmological simulations that the physical conditions in the accreting plasma can change markedly at the virial shock:  the gas transitions from predominantly photoionized at $\sim 10^4$ K to much hotter and collisionally ionized.   It is very likely that the magnetic field is significantly modified as well, with post-shock compression and turbulence amplifying the magnetic field.   As a result, the virial shock may be a location where CR transport changes significantly (in addition to being a source of shock-accelerated CRs; \citealt{Pinzke2010}).  The impact of this on CR feedback in massive halos will require detailed study that includes the dependence of CR transport on the local plasma conditions.

\section{Non-thermal Emission From BH-Generated CRs in Halo Outskirts}
\label{sec:NT}

The most direct confirmation of CR feedback in halo outskirts would be the detection of non-thermal emission at radii $\sim R_{200}$; such a detection could be used to measure the CR pressure.  Here we assess the feasibility of such observations, focusing on non-thermal emission produced by CR protons via pion decay, which produces gamma-rays, electron-positron pairs, and neutrinos.   CR electron cooling times for electrons that radiate in observable bands are short in the centers of massive halos and thus observable CR electrons are less likely to be energetically significant many Gyr later in cluster outskirts (low energy CR electrons may be present in halo outskirts and dynamically important [see \S \ref{sec:disc}] but they only radiate, e.g., synchrotron at frequencies $\ll 100$ MHz). 

There is an extensive body of observational  and theoretical work on non-thermal emission in galaxy clusters, both gamma-ray (e.g., \citealt{Ackerman2014}) and radio (e.g., \citealt{vanWeeren2019}).   Our estimates in this section differ in focusing on lower mass systems and on emission from CRs produced by BHs that have diffused to larger radii rather than, e.g., CRs produced by structure formation shocks.

The timescale for a CR proton to lose energy via pion decay is $t_{pion} \simeq 10^{12} \delta_{10}^{-1} \, {\rm yrs}$ where we have parameterized the local gas density near the virial radius in terms of the overdensity $\delta = 10 \, \delta_{10}$ relative to the critical density (i.e., $\rho_{\rm baryon} = \delta_{10} 10 \rho_{\rm crit}$) with $\delta \sim 10$ suggested by observations of cluster outskirts and simulations without strong feedback on scales of $\sim R_{200}$ \citep{Pratt2022}. The pion production cross-section (and thus $t_{pion}$) is nearly independent of CR proton energy a factor of few above the threshold energy of $\sim 0.28$ GeV.

We assume that the CR protons have an energy spectrum above a GeV of 
\begin{equation}
E \frac{dn}{d \ln E} = 3 (\alpha - 2) p_{CR} \left( \frac{E}{\rm GeV}\right)^{2-\alpha}
\label{eq:pspec}
\end{equation}
where $p_{CR}$ is the total CR pressure and $\alpha$ is the proton spectral index. We take $\alpha \simeq 2.2$ in what follows, which is a typical injection spectrum of CRs for the strong shocks generated by BHs (the predicted emission below is even fainter if we use a steeper proton spectrum, e.g., the local interstellar spectrum in the Milky Way).
In pion production, of the initial CR proton energy, 1/3 goes into goes into neutral pions and gamma-rays and the remaining 2/3 goes into charged pions and eventually pairs and neutrinos; the pairs receive $\simeq 1/4$ of the charged pion energy and thus $\simeq 1/6$ of the initial proton energy \citep{Schlickeiser2002}.   

The volumetric emissivity of gamma-rays of energy $E_\gamma$ produced by pion decay can be thus be estimated as $E_\gamma j_{E_\gamma} \simeq E dn/d\ln E |_{6 E_\gamma}/(3 t_{pion})$. 
The specific intensity of diffuse gamma-ray emission from pion decay in cluster outskirts is then roughly $E_\gamma j_{E_\gamma} R_{200}/(4 \pi)$, i.e.,
\begin{equation}
\begin{aligned}
& E_{\gamma} I_{E_\gamma}|_{r \sim R_{200}}  \simeq  \, 5 \times 10^{-12} \, {\rm erg \, cm^{-2} \, s^{-1} \, sr^{-1}}  \, \\ &  \ \ \ \ \ \ \ \ \ \times \delta_{10} \, \left(\frac{E_\gamma}{\rm GeV}\right)^{-0.2}  \, \left(\frac{p_{CR}}{p_{th,SS}}\right) \left(\frac{M_{200}}{10^{13} M_{\odot}}\right)
\end{aligned}
\label{eq:gintensity}
\end{equation}
In equation \ref{eq:gintensity} we have normalized the CR pressure to the self-similar gas pressure at $r = R_{200}$ from \citet{Arnaud2010},  $p_{th,SS}|_{R_{200}} \simeq 8 \times 10^{-15} \, {\rm erg \, cm^{-3}} (M_{200}/10^{13} M_\odot)^{2/3}$, motivated by the goal of assessing whether dynamically important CR pressure in halo outskirts produces observable non-thermal emission.

For comparison to equation \ref{eq:gintensity}, the extragalactic gamma-ray background intensity is $E_\gamma I_{E_{\gamma}} \simeq 10^{-9} (E_{\gamma}/{\rm GeV})^{-0.3} \, {\rm erg \, cm^{-2} \, s^{-1} \, sr^{-1}}$ \citep{gammaraybg}.  The intensity produced by dynamically important CR pressure at $r \sim R_{200}$ is well below that of the extragalactic gamma-ray background in group-mass systems.   Thus direct detection of non-thermal gamma-ray emission in halo outskirts appears unlikely.  A more promising possibility is detection of gamma-rays produced at smaller radii in high mass groups where the gas density is larger (ie., larger $\delta $ in eq. \ref{eq:gintensity}).   This may be feasible with improved gamma-ray telescopes such as the Cherenkov Telescope Array \citep{CTA} and would provide important new constraints on the CR content of massive halos.

A second possible observational signature of pion decay in massive halos is diffuse radio emission from secondary electron-positron pairs produced in charged pion decay.   The total energy injection rate into charged pairs $\dot \epsilon_e$ (each of which has energy $E_e$) via pion production by a proton of energy $E$ is given by
\begin{equation}
\dot \epsilon_e(E_e) \simeq \frac{E dn/d\ln E}{6 t_{pion}}
\label{eq:edote}
\end{equation}
where $E_e \simeq E/12$ because the pairs collectively receive 1/6 of the initial proton energy.   The dominant energy loss mechanism at large radii in halos is likely inverse Compton (IC) radiation off of the CMB (unless the CR transport timescale is much shorter than the IC cooling timescale, which will only further decrease the synchrotron luminosity estimated here).   The synchrotron specific intensity is then given by
\begin{equation}
\begin{aligned}
\nu I_\nu|_{r \sim R_{200}}  & \simeq  \frac{{E dn/d\ln E}|_{12 E_e}}{48 \pi t_{pion}} \frac{U_B}{U_{CMB}} R_{200}  \\
\, &  \sim 10^{-14} \, {\rm erg \, cm^{-2} \, s^{-1} \, sr^{-1}} \,  \nu_{0.1 GHz}^{-0.1} \frac{\delta_{10}}{\beta^{1.05}}   \\ &  \ \ \ \ \ \ \ \ \ \times \, \left(\frac{p_{CR}}{p_{th,SS}}\right) \left(\frac{M_{200}}{10^{13} M_{\odot}}\right)^{1.7}
\label{eq:rintensity}
\end{aligned}
\end{equation}
where an electron/positron of energy $E_e \simeq 3 \, {\rm GeV} \, B_{\mu G}^{-1/2} \nu_{0.1 GHz}^{1/2}$ dominates the synchrotron emission at frequency $\nu = 0.1 \nu_{0.1 GHz}$ GHz, $U_B$ is the magnetic energy density, $U_{CMB}$ is the energy density of the CMB, and in the second expression we have taken $B^2/4 \pi = P_{th,SS}/\beta$ as a convenient scaling of the magnetic field at $r \sim R_{200}$, with $\beta$ likely $\sim 100$ or larger.  For comparison to equation \ref{eq:rintensity}, the extragalactic radio background intensity at 100 MHz is $\nu I_\nu \sim 10^{-10}  \, {\rm erg \, cm^{-2} \, s^{-1} \, sr^{-1}}$ \citep{Tompkins2023}.  Given that $\beta$ is likely $\sim 100$, synchrotron emission from secondaries produced near $r \sim R_{200}$ with $p_{CR} \sim p_{th,SS}$ is far too faint to be observable even in $10^{14} M_\odot$ halos. 

Finally, we assess whether IC-scattered CMB photons from secondary pairs could outshine thermal free-free emission in cluster outskirts.   Since the IC cooling time at the energies of interest is shorter than a Hubble time, it is easy to show that the ratio of the IC to free-free emissivities in the X-rays is
\begin{equation}
\frac{j_{IC}}{j_{ff}} \sim 10^{-2} \left(\frac{p_{CR}}{p_{th}}\right) \left(\frac{T}{10^7 \, {\rm K}}\right)^{3/2}  \frac{\exp(E_X/kT)}{(E_X/{\rm keV})^{1.2}}
\label{eq:ICvsff}
\end{equation}
where $E_X$ is the photon energy.  Equation \ref{eq:ICvsff} shows that IC emission on the CMB from secondary pairs can thus only outshine the thermal emission for photon energies $E_X \gtrsim 5-10 kT$ where the thermal emission plummets.    

Taken together the estimates in this section show that it is unfortunately challenging to directly detect CR feedback via nonthermal emission at radii of order the virial radius in massive halos.   Deeper searches in high energy gamma-rays from the inner regions of groups (e.g., with the CTA) appears the most promising avenue for future work.   Deep radio and high energy X-ray studies of groups would also be valuable; if the transport time of primary CR electrons is shorter than their cooling time (the opposite of what we have assumed here), the radio and IC emission from primary CR electrons could be significantly larger than the emission from secondaries estimated here.

\section{Summary and Discussion}

\label{sec:disc}

In this paper we have argued that CRs produced by BH accretion during the primary epoch of BH growth at $z \sim 1-3$ may be a key source of feedback in massive halos.  We show that at lower redshifts CR feedback is likely to preferentially act in the outskirts of $\sim 10^{13-14} M_\odot$ halos, but not in the outskirts of yet more massive galaxy clusters (e.g., Fig. \ref{fig:CRR200} and \ref{fig:CRradial}).  More specifically, the CR pressure can be comparable to or exceed the thermal gas pressure at radii of order the virial radius.  This is primarily because the CR pressure declines  more slowly with increasing radius than the thermal gas pressure (Fig. \ref{fig:CRradial}).  CRs that are currently dynamically unimportant in the core of the halo can be dynamically dominant in the halo outskirts.  We suggest that the large CR pressure in the outskirts of group-mass halos is likely to drive baryons well outside the halo (by analogy, e.g., to models of CR-driven winds from galaxies; see \citealt{Thompson2024} for a review).  
An important feature of this feedback mechanism is that the properties of CR feedback in group-mass halos at large radii are not directly related to the current radiative cooling rate of hot gas at smaller radii, which is typically used as a proxy for `current' BH feedback.  In essence, this just reflects the well-established fact that most BH mass accretion happened early in massive halos, with relatively little present-day growth (e.g., \citealt{Shen_qlf2020}).   Since CRs in the outer halo have transport timescales of $\gtrsim$\,Gyr their properties are sensitive to the entire galaxy history, not the current BH activity.  

Our result that $p_{CR} \sim p_{th}$ in halo outskirts strongly suggests that BH-generated CRs are capable of driving gas from large halo radii out past the virial radius in $\sim 10^{13-14} M_\odot$ halos, as in the simulations of lower mass halos with SNe-generated CRs in Fig. \ref{fig:CGMCR}.  
We propose that CRs may thus be a key mechanism modifying the baryon distribution in group mass systems, as suggested by the weak lensing $S_8$ tension \citep{Preston2023} and recent measurements of the kinetic Sunvaev-Zeldovich effect \citep{Boryana2024,Bigwood2024}.  A significant uncertainty in our results is that the physics of CR transport is not fully understood theoretically.  This in turn affects the radial distribution of CRs in and around massive halos.   We have shown, however, that order of magnitude estimates of CR transport mediated by the streaming instability predict CR streaming speeds and effective diffusion coefficients of order those that lead to CR pressure exceeding thermal gas pressure in the outskirts of $\sim 10^{13-14} M_\odot$ halos (eq. \ref{eq:kappastr} and \S \ref{sec:transport}).  This strongly motivates more detailed exploration of the CR feedback mechanism proposed here.  

{A large dynamically important CR pressure in halo outskirts at lower redshift is likely associated with a large CR pressure in the inner halo at higher redshift when the central BH was active (see \S \ref{sec:CRfeedback}).  This conclusion can be avoided if, for example: (1) CR transport is much more rapid in the inner halo than outer halo; (2) the bulk of the CR energy is produced over sufficiently long timescales ($\gtrsim 10^{8}$ yrs) that the CRs diffuse to larger radii before their pressure builds to a dynamically significant level in the inner halo; or (3) CRs are sourced on sufficiently large length scales ($\gtrsim 0.1 R_{200}$) in extended jets.  Future cosmological simulations with BHs and CRs can explicitly address the latter possibilities.   Such simulations will  be critical for assessing whether models with strong CR feedback in the outer halo can produce groups whose inner halo properties  remains consistent with X-ray observations (even absent CRs, spatially extended injection of mechanical AGN feedback can help reconcile X-ray, SZ, and weak lensing observations; \citealt{Bigwood2025}).   One key may be that if most BH growth occurs when the halo baryons are accreting via cold filaments then, even if the CR pressure in the inner halo was higher at these times, the majority of the CRs will not impact the inner halo because of the much higher ram pressure associated with dense filamentary gas accretion.  The dynamical impact of  CRs will only become important once the halo fully virializes, by which point the CRs produced during the earlier BH growth phase will have diffused to larger radii.   Indeed, in the FIRE simulations of a MW-mass galaxy shown in Figure \ref{fig:CGMCR}, the impact of CRs is small at high redshift when most of the stars formed and most of the CRs were produced, in part for this reason  -- only at later times when the halo virializes do the CGM properties change dramatically as shown in the Figure.  A key difference between MW-mass and group-mass halos is that in the former, ongoing star formation produces enough CRs via core-collapse supernovae to modify the inner halo at late times post virialization.\footnote{The CR pressure profile in Figure \ref{fig:CGMCR} is quite different from the models in Figure \ref{fig:CRradial}.  It is instead similar to equation \ref{eq:PCREdot} because there is a continued source of CRs at low redshift from core-collapse SNe.  The dominant transport in the halo is streaming and advection with the gas so that the effective $\kappa \propto r$ and $p_{CR} \propto r^{-2}$.  See, e.g., \citet{Hopkins2021} for more discussion of the CR profiles in the CGM in these simulations.} In group mass halos, neither star formation nor ongoing BH activity likely do,  so the CR feedback may be dominated by where the majority of CRs produced by early BH growth reside -- in the outer halo.}

{It is also worth stressing that existing calibrations of outer halo feedback using galaxy group gas fractions and X-ray luminosities (e.g., \citealt{vanDaalen2020}) do not include models in which the dominant feedback in halo outskirts is by a separate CR fluid (as we are proposing here).  It is not at all clear that models with CR feedback will follow previously established correlations \citep{vanDaalen2020} between the matter power spectrum at $k \sim 1 \, {\rm Mpc^{-1}}$ and galaxy group gas fractions.  CR feedback is fundamentally different, in that the dynamical component responsible for the feedback is long-lived and active well after the mechanical feedback produced by BHs (e.g., jets, winds, `bubbles') has turned off.}

Our estimates of the CR pressure in halo outskirts explicitly neglect any loss of CR energy as the CRs travel from their sources to large halo radii.   This is a reasonable assumption for CR protons of almost all energies (except $\ll $ GeV) because pion loss timescales are long in diffuse halo plasmas. Moreover, AGN activity will inevitably produce a significant hadronic component via shocks produced by jets and winds in the ambient ISM and CGM/ICM.  However, the energy in CRs supplied via these shocks may be subdominant to the CR power directly supplied by relativistic jets.   It is possible that jets are primarily an electron-positron plasma rather than an electron-proton plasma (e.g., \citealt{blandford:2019.jet.review}).   In this case radiative losses are much more significant and can affect the total CR pressure at large halo radii.  In particular, only leptons with energies of $\sim 0.1-1$ GeV have energy loss timescales $\gtrsim$ a Gyr in diffuse halo plasmas (at the lower end of this energy range, the dominant loss is via Coulomb collisions, while at the higher energy end it is via synchrotron and Inverse Compton radiation).   However, these lower energy leptons are  those that typically dominate the total lepton CR pressure and so it is  plausible that even in leptonic models the CR pressure at large halo radii would be of order that estimated here.

There are two particularly direct tests of the role of CR feedback in halo outskirts.   The first is a comparison of Sunyaev-Zeldovich or X-ray inferred pressures relative to the hydrostatic pressure expected given the dark matter halo properties.  Any discrepancy would point to a non-thermal pressure source.   This measurement is challenging because it requires accurate gas density and pressure measurements at large radii in group-mass halos.    The detection of diffuse nonthermal CR emission at radii of order the virial radius or larger would also provide direct evidence for non-thermal pressure, specifically in the form of CRs.   Unfortunately, we  showed in \S \ref{sec:NT} that even if $p_{CR} \sim p_{th}$ in halo outskirts, nonthermal CR emission from CR protons and the secondary electron-positron pairs they produce is far too dim to detect in either the radio (synchrotron), X-ray (Inverse Compton) or gamma-rays (pion decay).   If GeV primary electrons are indeed energetically important at large halo radii they can produce significant Inverse Compton X-ray emission by upscattering CMB photons.  In \citet{Hopkins2025}, we show that this is likely the dominant source of extended soft X-ray emission in Milky Way and Andromeda mass halos, providing strong support for the importance of CRs in halo gas.   However, the Inverse Compton contribution to X-ray emission is subdominant to thermal emission in the more massive halos focused on in this paper.

Further study of the Inverse Compton X-ray halos produced by CR electrons in MW and M31-mass systems will provide critical constraints on CR propagation in the CGM that can inform models of CR feedback in more massive halos.   In addition, future high energy gamma-ray observations of pion decay with facilities like the Cherenkov Telescope Array \citep{CTA} will better constrain CRs in group mass halos.   These observations would not directly probe the CR content at the virial radius, but rather at smaller radii where most of the baryonic mass resides.   Existing Fermi observations rule out $p_{CR} \gtrsim p_{th}$ at the virial radius in massive galaxy clusters \citep{Ackerman2014} but the constraints are far less stringent in groups because the gamma-ray luminosity scales roughly linearly with the baryonic mass of the halo.   Moreover, we showed in \S \ref{sec:CRfeedback} that there is a large parameter space of CR diffusion coefficients in which CRs dominate the pressure in the outskirts of groups but satisfy the Fermi upper limits on the CR pressure in massive clusters (Fig. \ref{fig:CRradial}).    

In addition to driving baryons out of the halo via CR pressure, the presence of $p_{CR} \sim p_{th}$ at large halo radii might produce significant {CR-heating} of the thermal plasma.  
Streaming CRs inevitably heat the ambient plasma on a timescale (see Appendix \ref{sec:app}) $\sim (p_{th}/p_{CR}) \beta^{1/2} t_{dyn}$ where $\beta$ is the ratio of gas to magnetic pressure and $t_{dyn}$ is the local dynamical time (this heating is specific to CR transport via the streaming instability; it does not occur for pure diffusive CR transport).  If $\beta \lesssim 100$ and $p_{CR} \sim p_{th}$ this heating timescale could be of order the Hubble time.   In addition, protons and leptons with energies $\ll 1$\,GeV lose energy rapidly via Coulomb collisions with the ambient gas. This could be a non-negligible contribution to thermal balance at large radii if $p_{CR} \gtrsim p_{th}$ (and if low energy CRs are transported to large radii without losing most of their energy enroute).    Modifications of the thermal profile at large halo radii in addition to the density profile  may be detectable by future Sunyaev-Zeldovich measurements.

An interesting question that merits more study is the physics of CR transport as black hole-generated CRs reach the virial shock at large radii.   The virial shock represents a large change in the thermodynamic properties of the plasma, with the temperature, thermal pressure, and magnetic field strength all likely increasing signficantly across the virial shock.    In cosmological simulations of MW-mass halos with CRs produced by supernovae, CR-driven outflows can suppress the virial shock by pushing the halo baryons to significantly larger radii (see Fig. \ref{fig:CGMCR}; \citealt{Ji2021}).  These calculations assumed a constant diffusion coefficient, but in reality CR transport is likely to change significantly at the virial shock.   One possibility is that  slower transport exterior to the virial shock (because of weaker magnetic fields) leads to a build up of CR pressure just interior to the shock, further contributing to CRs driving the baryons and the virial shock to larger radii.

{For a fixed fraction of the energy produced by BHs going into CRs, simple energetics predict that the impact of BHs is likely to be the strongest in $\lesssim 10^{12} M_\odot$ halos (Fig. \ref{fig:CRtot}).   Indeed, some cosmological zoom-in simulations that incorporate CR feedback find that the same dimensionless feedback parameters (e.g., analogues of our $\epsilon_{\rm BH}$; eq. \ref{eq:EtotCRBH}) that can produce quenching in massive galaxies tend to overproduce feedback in lower mass halos \citep{Byrne2024}.  This may imply that the efficiency of jet and CR production is itself a function of BH mass, an effect that is not included in our estimates in Figure \ref{fig:CRR200}.   There is observational evidence that in the local Universe, radio emission from jets is dominated by the most massive BHs  \citep{Best2005}. It is unclear, however, if this applies at redshifts $\sim 2-3$ where we have argued the bulk of CR energy is likely produced.   Extensions of the work presented here accounting for possible BH-mass dependence of CR feedback would be useful for assessing the relative importance of SNe and BH CR feedback in $\sim 10^{12}$ halos that host MW-like galaxies. It is plausible, e.g., that BHs and SNe are similarly important for CR production in $10^{12} M_\odot$ halos, with BHs becoming increasingly dominant at higher halo masses.} 

A natural next step in evaluating the role of CR feedback in groups is to carry out simulations of massive halos with black hole-generated CRs (as in, e.g., \citealt{Wellons2023,Byrne2024}, whose zoom-in simulations included CRs from black holes in $\lesssim 10^{13} M_\odot$ halos).   The analytic estimates presented here motivate the CR energetics and transport properties to explore in such simulations.   The numerical tools for studying CR feedback in idealized and cosmological simulations exist for both diffusive and streaming transport (e.g., \citealt{Butsky2018, Jiang2018,Thomas2019,Hopkins2020a,Ramesh2024}).   Such simulations would address many key questions that we can only pose here:   Does CR feedback indeed produce winds at radii of order the virial radius in $10^{13-14} M_\odot$ halos and do such winds drive baryons to well outside the virialized halo?  What is the relative role of CRs generated by the central galaxy vs.  satellites (the satellites have $
\sim$ twice the stellar and black hole mass of the central in $\sim 10^{14} M_\odot$ halos)?  The same simulations will also enable quantitative observational tests of the feedback mechanism proposed here:   Do models with strong CR feedback in halo outskirts predict a pion gamma-ray flux detectable with CTAs in nearby groups?  Do models with strong CR feedback in group-mass systems produce galaxy clusters (which assemble from groups!) consistent with observations (a sort of `CR-preheating' question)?  Can CR feedback in groups indeed modify the dark matter distribution and kSZ signal at large radii while remaining consistent with X-ray constraints on the baryon fraction and gas profiles at smaller radii (as we have speculated may occur in this paper)?   Does it do so equally effectively in the halos that dominate the weak lensing and kSZ signals?   Fortunately the tools are in hand, guided by the estimates in this paper, to address these  questions and sharpen our understanding of the impact of CRs on structure formation.

\begin{acknowledgments}
EQ thanks Alex Amon for many conversations that stimulated this work and for valuable comments on an initial draft of the paper.  We also thank Philipp Kempski and Yan-Fei Jiang for useful conversations and the referee for useful feedback.   This work was supported in part by NSF AST grant 2107872 and by Simons Investigator grants to EQ and PFH.  PFH thanks the Institute for Advanced Study for sabbatical support. 
\end{acknowledgments}

\vspace{5mm}
    
\appendix

\section{Cosmic-ray Streaming}

\label{sec:app}

In this Appendix we describe some aspects of CR streaming and how it differs from diffusion.  We focus on   CR heating of the gas and its affect on the total energy content of the CRs, since this is  particularly relevant for the results in the main text (for more comprehensive discussions of CR transport, see the reviews by \citealt{Zweibel2017} and \citealt{Ruszkowski2023}).  For the interested reader not versed in the distinction between streaming and diffusive transport, we also compare a simple solution of the time dependent CR equations with streaming and diffusive transport.  Throughout this Appendix we intentionally consider a very simple model in which we hold the gas properties and magnetic field fixed in time, neglect the velocity of the gas, neglect catastrophic (e.g., pionic) and radiative CR losses, and treat the CRs as a single ultra-relativistic fluid whose dynamics we model on scales above the scattering mean-free-path, so that a fluid description is reasonable.  

The equation for the CR energy density $E_c$ in the absence of CR sources, and including transport by streaming at the Alfv\'en velocity down the CR pressure gradient and diffusion along magnetic field lines, is then given by
(e.g., \citealt{Zweibel2017})
\begin{equation}
\frac{\partial E_c}{\partial t}+\nabla\cdot {\bf F}_c={\bf v_{\rm s}}\cdot \nabla p_{c},
\label{cr_energy}
\end{equation}
where $p_c=E_c/3$ is the CR pressure,  ${\bf v_s} = -{\bf v_A}|\nabla p_c|/\nabla p_c$, ${\bf v_A}={\bf B}/(4\pi\rho)^{1/2}$, and ${\bf F_c} = 4 {\bf v_s} p_c -\kappa \,{\bf n}\left({\bf n}\cdot\nabla E_c\right)$; here $\kappa$ is the diffusion coefficient and ${\bf n}={\bf v_{\rm A}}/|\bf v_{\rm A}|$.   Note that the right hand side of equation \ref{cr_energy} is negative definite.  This is because CRs lose energy to heating the thermal gas via streaming-generated waves \citep{Wentzel1971}.  

In steady state, neglecting CR diffusion and assuming a well-defined sign of ${\bf v_s}$ (i.e., $\nabla p_c \ne 0$) one can rewrite eq. \ref{cr_energy} using ${\bf \nabla \cdot B} = 0$  as
\begin{equation}
4 \, {\bf B \cdot \nabla} \left(\frac{p_c}{\sqrt{\rho}}\right) = \frac{\bf B \cdot \nabla p_c}{\sqrt{\rho}}
\label{eq:pcr2}
\end{equation}
Equation \ref{cr_energy2}
gives the usual result that in streaming transport in steady state with fluid velocities $v \ll v_A$, $p_c \propto \rho^{2/3}$ along field lines (assuming $\rho$ and $p_c$ are decreasing).  What does this imply for the CR power on a flux tube of area A: $\dot E = 4 A p_c v_{A}$?   The CR power should decrease along the flux tube as CR energy is lost to the gas via the $v_A \cdot \nabla p_c$ heating term.  Note that this is different than for diffusive transport in which the CR power is conserved in the absence of PdV work on the gas (which we are not considering here because we are assuming the gas properties are fixed).    Using $A B \sim$const from magnetic flux conservation and $p_c \propto \rho^{2/3}$, it follows that $\dot E \propto \rho^{1/6}$.  This result is the standard result for how the CR streaming power changes with radius/density in galactic wind models, which consider individual flux tubes or a split-monopole geometry \citep{Ipavich1975,Mao2018,Quataert2022}.  In galaxy halos the magnetic field geometry is likely much more complex but along a given flux tube $\dot E \propto \rho^{1/6}$ holds in steady state so long as the density is decreasing.   As CRs stream out radially (on average) down the CR pressure gradient they will move from field line to field line stochastically and on average $\dot E \propto \rho^{1/6}$ will describe the {\rm radial} variation of the CR power as the CRs travel out in the halo.  One subtlety is that some field lines will inevitably have CRs attempting to stream up the density gradient, i.e., with increasing $\rho$.  However, if the density increases along a flux tube, the streaming solutions $p_c \propto \rho^{2/3}$ and $\dot E \propto \rho^{1/6}$ are inconsistent because the CRs can only stream down their pressure gradient not up it.   In regions where $\rho$ increases along a flux tube, the only physical solution (neglecting CR diffusion) is that the CR pressure is flat, a region called a CR `bottleneck' \citep{Wiener2017}.  The absence of a CR pressure gradient  implies that there is no driving of the streaming instability and hence no scattering to pin the CR streaming speed at $v_A$ in regions where the density increases along a flux tube.  In simulations of galactic winds, \citet{Quataert2022} found that the presence of bottlenecks on average {decreases} the amount of CR heating of the gas, i.e., leads to $\dot E \propto \rho^{\zeta}$ with $0 < \zeta < 1/6$.  

Consider now the time-dependent problem focused on in the main text.   If we again neglect diffusion and consider a region with a well-defined streaming direction so that ${\bf v_s}$ is not changing sign, using ${\bf \nabla \cdot B = 0}$, 
equation \ref{cr_energy} can be rewritten as
\begin{equation}
\frac{\partial p_c}{\partial t}+\nabla\cdot ({\bf v_s} p_c)= -\frac{p_c}{6}|{\bf v_{\rm A}}\cdot \nabla \ln \rho|
\label{cr_energy2}
\end{equation}
Equation \ref{cr_energy2} shows that for the time dependent problem the characteristic timescale scale over which the CR energy will decrease due to streaming losses is roughly 6 Alfv\'en crossing times of a density scale-height.

In the main text we neglected energy transfer from the CRs to the gas by using the solution of the diffusion equation to describe the spatial distribution of the CRs.  We further argued that the identification $\kappa \sim r v_A$ can transform between diffusion and streaming solutions.   This is only approximate, of course, in part because with CR streaming there is inevitably some energy transfer from the CRs to the gas that is not present in diffusive transport.  However, because the characteristic length scale for the CR energy to change significantly due to streaming losses is many density scale-heights, this change in energy is modest for our problem, perhaps a factor of $\sim 3$ out to the virial radius; this is a smaller effect than the overall uncertainties in the correct CR diffusion coefficient, total energy in CRs from BH growth, the magnetic field strength in halo outskirts, adiabatic or turbulent reacceleration, etc.

\begin{figure}[ht]
\centering
\includegraphics[width=0.5 \linewidth]{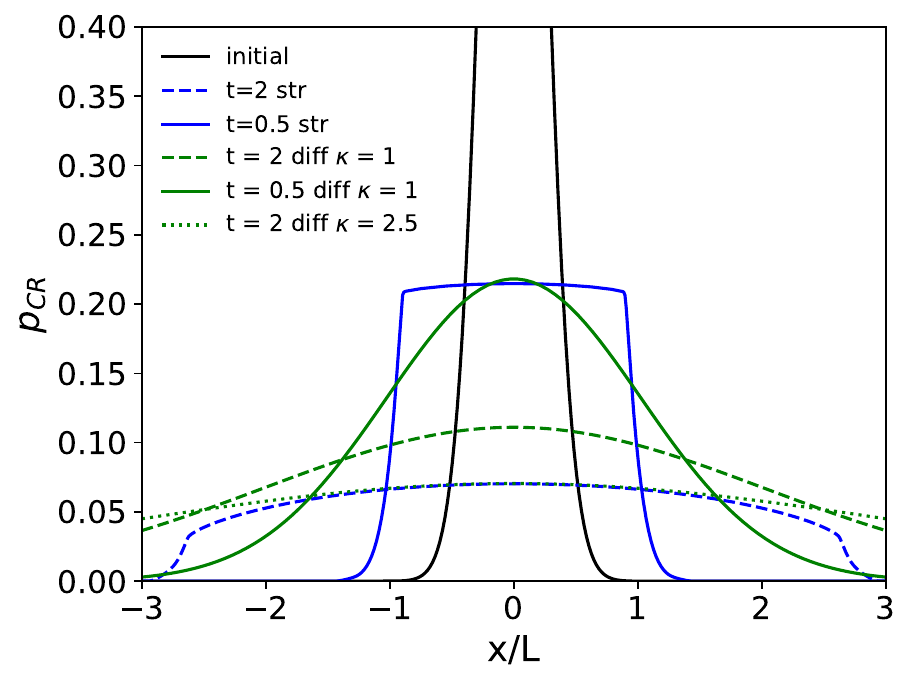}
\caption{Comparison of streaming vs. diffusive CR transport in a simple 1D problem.  The initial condition is a Gaussian in CR pressure centered at $x = 0$. The Alfv\'en velocity $v_A = 1$ and is assumed constant.  The gas properties are not evolved.  Diffusion coefficients are in units of $v_A L$ and time is in units of $L/v_A$.   Although the solutions differ in detail both streaming and diffusive transport produce CR pressure profiles that broaden significantly over time in a qualitatively similar manner.     
\label{fig:strvsdiff}}
\end{figure}

Figure \ref{fig:strvsdiff} compares solutions of the one-dimensional time-dependent CR equations with diffusive and streaming transport.   The gas properties are assumed fixed with a constant Alfv\'en speed $v_A = 1$, so only the CR pressure and flux are evolved.  The initial condition is a Gaussian centered at $x = 0$.   With streaming transport the Gaussian widens into a top-hat like profile that broadens in time.   The diffusion equation satisfies the usual time dependent solution $p_c(x,t) = (4 \kappa \alpha t + 1)^{-1/2} \exp[-\alpha x^2/(4 \kappa \alpha t + 1)]$ where $\alpha = 10$ is set by the initial condition.   Although the streaming and diffusive models differ in detail, they are qualitatively similar in producing profiles that spread out over time reaching a roughly constant value over a length scale $\sim v_A t$ (streaming) or $\sim \sqrt{\kappa t}$ (diffusion).   In particular, although  streaming CRs resembles advection at the Alfv\'en speed, because it is streaming down the CR pressure gradient (which changes sign in the initial condition here), streaming leads to a broadening of the initial CR distribution analogous to that produced by diffusion.    Note also that at later times, the CR diffusion coefficient that better qualitatively matches the streaming solution increases, broadly consistent with the identification $\kappa \sim r v_A$ in the main text.   Finally, it is interesting to note that for this specific test problem equation \ref{cr_energy2} does not apply.   In particular, the right hand side is zero since we assume constant $B, \rho,$ and $v_A$.  There is nonetheless some loss of CR energy in the streaming solution associated with where ${\bf v_s}$ changes sign and so the derivation of equation \ref{eq:pcr2} does not hold.   The total CR energy decreases by $\sim 50\%$ by $t = 2$ in the streaming solution, while it is of course constant for the diffusive solution.

\bibliographystyle{mn2e}
\bibliography{main}

\begin{thebibliography}{}
\makeatletter
\relax
\def\mn@urlcharsother{\let\do\@makeother \do\$\do\&\do\#\do\^\do\_\do\%\do\~}
\def\mn@doi{\begingroup\mn@urlcharsother \@ifnextchar [ {\mn@doi@} {\mn@doi@[]}}
\def\mn@doi@[#1]#2{\def\@tempa{#1}\ifx\@tempa\@empty \href {http://dx.doi.org/#2} {doi:#2}\else \href {http://dx.doi.org/#2} {#1}\fi \endgroup}
\def\mn@eprint#1#2{\mn@eprint@#1:#2::\@nil}
\def\mn@eprint@arXiv#1{\href {http://arxiv.org/abs/#1} {{\tt arXiv:#1}}}
\def\mn@eprint@dblp#1{\href {http://dblp.uni-trier.de/rec/bibtex/#1.xml} {dblp:#1}}
\def\mn@eprint@#1:#2:#3:#4\@nil{\def\@tempa {#1}\def\@tempb {#2}\def\@tempc {#3}\ifx \@tempc \@empty \let \@tempc \@tempb \let \@tempb \@tempa \fi \ifx \@tempb \@empty \def\@tempb {arXiv}\fi \@ifundefined {mn@eprint@\@tempb}{\@tempb:\@tempc}{\expandafter \expandafter \csname mn@eprint@\@tempb\endcsname \expandafter{\@tempc}}}

\bibitem[\protect\citeauthoryear{{Ackermann} et~al.,}{{Ackermann} et~al.}{2014}]{Ackerman2014}
{Ackermann} M.,  et~al., 2014, \mn@doi [\apj] {10.1088/0004-637X/787/1/18}, \href {https://ui.adsabs.harvard.edu/abs/2014ApJ...787...18A} {787, 18}

\bibitem[\protect\citeauthoryear{{Ackermann} et~al.,}{{Ackermann} et~al.}{2015}]{gammaraybg}
{Ackermann} M.,  et~al., 2015, \mn@doi [\apj] {10.1088/0004-637X/799/1/86}, \href {https://ui.adsabs.harvard.edu/abs/2015ApJ...799...86A} {799, 86}

\bibitem[\protect\citeauthoryear{{Adam}, {Goksu}, {Brown}, {Rudnick}  \& {Ferrari}}{{Adam} et~al.}{2021}]{Adam2021}
{Adam} R.,  {Goksu} H.,  {Brown} S.,  {Rudnick} L.,   {Ferrari} C.,  2021, \mn@doi [\aap] {10.1051/0004-6361/202039660}, \href {https://ui.adsabs.harvard.edu/abs/2021A&A...648A..60A} {648, A60}

\bibitem[\protect\citeauthoryear{{Amon} \& {Efstathiou}}{{Amon} \& {Efstathiou}}{2022}]{Amon2022}
{Amon} A.,  {Efstathiou} G.,  2022, \mn@doi [\mnras] {10.1093/mnras/stac2429}, \href {https://ui.adsabs.harvard.edu/abs/2022MNRAS.516.5355A} {516, 5355}

\bibitem[\protect\citeauthoryear{{Amon} et~al.,}{{Amon} et~al.}{2022}]{Amon2022b}
{Amon} A.,  et~al., 2022, \mn@doi [\prd] {10.1103/PhysRevD.105.023514}, \href {https://ui.adsabs.harvard.edu/abs/2022PhRvD.105b3514A} {105, 023514}

\bibitem[\protect\citeauthoryear{{Anderson}, {Gaspari}, {White}, {Wang}  \& {Dai}}{{Anderson} et~al.}{2015}]{Anderson2015}
{Anderson} M.~E.,  {Gaspari} M.,  {White} S. D.~M.,  {Wang} W.,   {Dai} X.,  2015, \mn@doi [\mnras] {10.1093/mnras/stv437}, \href {https://ui.adsabs.harvard.edu/abs/2015MNRAS.449.3806A} {449, 3806}

\bibitem[\protect\citeauthoryear{{Arnaud}, {Pratt}, {Piffaretti}, {B{\"o}hringer}, {Croston}  \& {Pointecouteau}}{{Arnaud} et~al.}{2010}]{Arnaud2010}
{Arnaud} M.,  {Pratt} G.~W.,  {Piffaretti} R.,  {B{\"o}hringer} H.,  {Croston} J.~H.,   {Pointecouteau} E.,  2010, \mn@doi [\aap] {10.1051/0004-6361/200913416}, \href {https://ui.adsabs.harvard.edu/abs/2010A&A...517A..92A} {517, A92}

\bibitem[\protect\citeauthoryear{{Asgari} et~al.,}{{Asgari} et~al.}{2021}]{Asgari2021}
{Asgari} M.,  et~al., 2021, \mn@doi [\aap] {10.1051/0004-6361/202039070}, \href {https://ui.adsabs.harvard.edu/abs/2021A&A...645A.104A} {645, A104}

\bibitem[\protect\citeauthoryear{{Baghmanyan}, {Zargaryan}, {Aharonian}, {Yang}, {Casanova}  \& {Mackey}}{{Baghmanyan} et~al.}{2022}]{Baghamanyan2022}
{Baghmanyan} V.,  {Zargaryan} D.,  {Aharonian} F.,  {Yang} R.,  {Casanova} S.,   {Mackey} J.,  2022, \mn@doi [\mnras] {10.1093/mnras/stac2266}, \href {https://ui.adsabs.harvard.edu/abs/2022MNRAS.516..562B} {516, 562}

\bibitem[\protect\citeauthoryear{{Bai}, {Ostriker}, {Plotnikov}  \& {Stone}}{{Bai} et~al.}{2019}]{Bai2019}
{Bai} X.-N.,  {Ostriker} E.~C.,  {Plotnikov} I.,   {Stone} J.~M.,  2019, \mn@doi [\apj] {10.3847/1538-4357/ab1648}, \href {https://ui.adsabs.harvard.edu/abs/2019ApJ...876...60B} {876, 60}

\bibitem[\protect\citeauthoryear{{Begelman} \& {Cioffi}}{{Begelman} \& {Cioffi}}{1989}]{Begelman1989}
{Begelman} M.~C.,  {Cioffi} D.~F.,  1989, \mn@doi [\apjl] {10.1086/185542}, \href {https://ui.adsabs.harvard.edu/abs/1989ApJ...345L..21B} {345, L21}

\bibitem[\protect\citeauthoryear{{Best}, {Kauffmann}, {Heckman}, {Brinchmann}, {Charlot}, {Ivezi{\'c}}  \& {White}}{{Best} et~al.}{2005}]{Best2005}
{Best} P.~N.,  {Kauffmann} G.,  {Heckman} T.~M.,  {Brinchmann} J.,  {Charlot} S.,  {Ivezi{\'c}} {\v{Z}}.,   {White} S.~D.~M.,  2005, \mn@doi [\mnras] {10.1111/j.1365-2966.2005.09192.x}, \href {https://ui.adsabs.harvard.edu/abs/2005MNRAS.362...25B} {362, 25}

\bibitem[\protect\citeauthoryear{{Bigwood} et~al.,}{{Bigwood} et~al.}{2024}]{Bigwood2024}
{Bigwood} L.,  et~al., 2024, \mn@doi [\mnras] {10.1093/mnras/stae2100}, \href {https://ui.adsabs.harvard.edu/abs/2024MNRAS.534..655B} {534, 655}

\bibitem[\protect\citeauthoryear{{Bigwood}, {Bourne}, {Irsic}, {Amon}  \& {Sijacki}}{{Bigwood} et~al.}{2025}]{Bigwood2025}
{Bigwood} L.,  {Bourne} M.~A.,  {Irsic} V.,  {Amon} A.,   {Sijacki} D.,  2025, \mn@doi [arXiv e-prints] {10.48550/arXiv.2501.16983}, \href {https://ui.adsabs.harvard.edu/abs/2025arXiv250116983B} {p. arXiv:2501.16983}

\bibitem[\protect\citeauthoryear{{B{\^\i}rzan}, {Rafferty}, {McNamara}, {Wise}  \& {Nulsen}}{{B{\^\i}rzan} et~al.}{2004}]{Birzan2004}
{B{\^\i}rzan} L.,  {Rafferty} D.~A.,  {McNamara} B.~R.,  {Wise} M.~W.,   {Nulsen} P.~E.~J.,  2004, \mn@doi [\apj] {10.1086/383519}, \href {https://ui.adsabs.harvard.edu/abs/2004ApJ...607..800B} {607, 800}

\bibitem[\protect\citeauthoryear{{Blandford}, {Meier}  \& {Readhead}}{{Blandford} et~al.}{2019}]{blandford:2019.jet.review}
{Blandford} R.,  {Meier} D.,   {Readhead} A.,  2019, \mn@doi [\araa] {10.1146/annurev-astro-081817-051948}, \href {https://ui.adsabs.harvard.edu/abs/2019ARA&A..57..467B} {57, 467}

\bibitem[\protect\citeauthoryear{{Butsky} \& {Quinn}}{{Butsky} \& {Quinn}}{2018}]{Butsky2018}
{Butsky} I.~S.,  {Quinn} T.~R.,  2018, \mn@doi [\apj] {10.3847/1538-4357/aaeac2}, \href {https://ui.adsabs.harvard.edu/abs/2018ApJ...868..108B} {868, 108}

\bibitem[\protect\citeauthoryear{{Butsky} et~al.,}{{Butsky} et~al.}{2022}]{Butsky2022}
{Butsky} I.~S.,  et~al., 2022, \mn@doi [\apj] {10.3847/1538-4357/ac7ebd}, \href {https://ui.adsabs.harvard.edu/abs/2022ApJ...935...69B} {935, 69}

\bibitem[\protect\citeauthoryear{{Butsky}, {Hopkins}, {Kempski}, {Ponnada}, {Quataert}  \& {Squire}}{{Butsky} et~al.}{2024}]{Butsky2024}
{Butsky} I.~S.,  {Hopkins} P.~F.,  {Kempski} P.,  {Ponnada} S.~B.,  {Quataert} E.,   {Squire} J.,  2024, \mn@doi [\mnras] {10.1093/mnras/stae276}, \href {https://ui.adsabs.harvard.edu/abs/2024MNRAS.528.4245B} {528, 4245}

\bibitem[\protect\citeauthoryear{{Byrne} et~al.,}{{Byrne} et~al.}{2024}]{Byrne2024}
{Byrne} L.,  et~al., 2024, \mn@doi [\apj] {10.3847/1538-4357/ad67ca}, \href {https://ui.adsabs.harvard.edu/abs/2024ApJ...973..149B} {973, 149}

\bibitem[\protect\citeauthoryear{{Chandran}}{{Chandran}}{2000}]{Chandran2000}
{Chandran} B. D.~G.,  2000, \mn@doi [\prl] {10.1103/PhysRevLett.85.4656}, \href {https://ui.adsabs.harvard.edu/abs/2000PhRvL..85.4656C} {85, 4656}

\bibitem[\protect\citeauthoryear{{Cherenkov Telescope Array Consortium} et~al.,}{{Cherenkov Telescope Array Consortium} et~al.}{2019}]{CTA}
{Cherenkov Telescope Array Consortium} et~al., 2019, {Science with the Cherenkov Telescope Array}, \mn@doi{10.1142/10986.
}

\bibitem[\protect\citeauthoryear{{Churazov}, {Forman}, {Vikhlinin}, {Tremaine}, {Gerhard}  \& {Jones}}{{Churazov} et~al.}{2008}]{Churazov2008}
{Churazov} E.,  {Forman} W.,  {Vikhlinin} A.,  {Tremaine} S.,  {Gerhard} O.,   {Jones} C.,  2008, \mn@doi [\mnras] {10.1111/j.1365-2966.2008.13507.x}, \href {https://ui.adsabs.harvard.edu/abs/2008MNRAS.388.1062C} {388, 1062}

\bibitem[\protect\citeauthoryear{{Connor} et~al.,}{{Connor} et~al.}{2024}]{Connor2024}
{Connor} L.,  et~al., 2024, \mn@doi [arXiv e-prints] {10.48550/arXiv.2409.16952}, \href {https://ui.adsabs.harvard.edu/abs/2024arXiv240916952C} {p. arXiv:2409.16952}

\bibitem[\protect\citeauthoryear{{Croton} et~al.,}{{Croton} et~al.}{2006}]{Croton2006}
{Croton} D.~J.,  et~al., 2006, \mn@doi [\mnras] {10.1111/j.1365-2966.2005.09675.x}, \href {https://ui.adsabs.harvard.edu/abs/2006MNRAS.365...11C} {365, 11}

\bibitem[\protect\citeauthoryear{{Fabian}}{{Fabian}}{2012}]{Fabian2012}
{Fabian} A.~C.,  2012, \mn@doi [\araa] {10.1146/annurev-astro-081811-125521}, \href {https://ui.adsabs.harvard.edu/abs/2012ARA&A..50..455F} {50, 455}

\bibitem[\protect\citeauthoryear{{Fu} et~al.,}{{Fu} et~al.}{2022}]{Fu2022}
{Fu} H.,  et~al., 2022, \mn@doi [\mnras] {10.1093/mnras/stac2205}, \href {https://ui.adsabs.harvard.edu/abs/2022MNRAS.516.3206F} {516, 3206}

\bibitem[\protect\citeauthoryear{{Guo} \& {Oh}}{{Guo} \& {Oh}}{2008}]{Guo2008}
{Guo} F.,  {Oh} S.~P.,  2008, \mn@doi [\mnras] {10.1111/j.1365-2966.2007.12692.x}, \href {https://ui.adsabs.harvard.edu/abs/2008MNRAS.384..251G} {384, 251}

\bibitem[\protect\citeauthoryear{{Hadzhiyska} et~al.,}{{Hadzhiyska} et~al.}{2024}]{Boryana2024}
{Hadzhiyska} B.,  et~al., 2024, \mn@doi [arXiv e-prints] {10.48550/arXiv.2407.07152}, \href {https://ui.adsabs.harvard.edu/abs/2024arXiv240707152H} {p. arXiv:2407.07152}

\bibitem[\protect\citeauthoryear{{Hopkins} et~al.,}{{Hopkins} et~al.}{2020}]{Hopkins2020a}
{Hopkins} P.~F.,  et~al., 2020, \mn@doi [\mnras] {10.1093/mnras/stz3321}, \href {https://ui.adsabs.harvard.edu/abs/2020MNRAS.492.3465H} {492, 3465}

\bibitem[\protect\citeauthoryear{{Hopkins}, {Chan}, {Ji}, {Hummels}, {Kere{\v{s}}}, {Quataert}  \& {Faucher-Gigu{\`e}re}}{{Hopkins} et~al.}{2021a}]{Hopkins2021}
{Hopkins} P.~F.,  {Chan} T.~K.,  {Ji} S.,  {Hummels} C.~B.,  {Kere{\v{s}}} D.,  {Quataert} E.,   {Faucher-Gigu{\`e}re} C.-A.,  2021a, \mn@doi [\mnras] {10.1093/mnras/staa3690}, \href {https://ui.adsabs.harvard.edu/abs/2021MNRAS.501.3640H} {501, 3640}

\bibitem[\protect\citeauthoryear{{Hopkins}, {Chan}, {Squire}, {Quataert}, {Ji}, {Kere{\v{s}}}  \& {Faucher-Gigu{\`e}re}}{{Hopkins} et~al.}{2021b}]{Hopkins2021c}
{Hopkins} P.~F.,  {Chan} T.~K.,  {Squire} J.,  {Quataert} E.,  {Ji} S.,  {Kere{\v{s}}} D.,   {Faucher-Gigu{\`e}re} C.-A.,  2021b, \mn@doi [\mnras] {10.1093/mnras/staa3692}, \href {https://ui.adsabs.harvard.edu/abs/2021MNRAS.501.3663H} {501, 3663}

\bibitem[\protect\citeauthoryear{{Hopkins}, {Quataert}, {Ponnada}  \& {Silich}}{{Hopkins} et~al.}{2025}]{Hopkins2025}
{Hopkins} P.~F.,  {Quataert} E.,  {Ponnada} S.~B.,   {Silich} E.,  2025, arXiv e-prints, \href {https://ui.adsabs.harvard.edu/abs/2025arXiv250118696H} {p. arXiv:2501.18696}

\bibitem[\protect\citeauthoryear{{Huber}, {Tchernin}, {Eckert}, {Farnier}, {Manalaysay}, {Straumann}  \& {Walter}}{{Huber} et~al.}{2013}]{Huber2013}
{Huber} B.,  {Tchernin} C.,  {Eckert} D.,  {Farnier} C.,  {Manalaysay} A.,  {Straumann} U.,   {Walter} R.,  2013, \mn@doi [\aap] {10.1051/0004-6361/201321947}, \href {https://ui.adsabs.harvard.edu/abs/2013A&A...560A..64H} {560, A64}

\bibitem[\protect\citeauthoryear{{Ipavich}}{{Ipavich}}{1975}]{Ipavich1975}
{Ipavich} F.~M.,  1975, \mn@doi [\apj] {10.1086/153397}, \href {http://adsabs.harvard.edu/abs/1975ApJ...196..107I} {196, 107}

\bibitem[\protect\citeauthoryear{{Ji} et~al.,}{{Ji} et~al.}{2020}]{Ji2020}
{Ji} S.,  et~al., 2020, \mn@doi [\mnras] {10.1093/mnras/staa1849}, \href {https://ui.adsabs.harvard.edu/abs/2020MNRAS.496.4221J} {496, 4221}

\bibitem[\protect\citeauthoryear{{Ji}, {Kere{\v{s}}}, {Chan}, {Stern}, {Hummels}, {Hopkins}, {Quataert}  \& {Faucher-Gigu{\`e}re}}{{Ji} et~al.}{2021}]{Ji2021}
{Ji} S.,  {Kere{\v{s}}} D.,  {Chan} T.~K.,  {Stern} J.,  {Hummels} C.~B.,  {Hopkins} P.~F.,  {Quataert} E.,   {Faucher-Gigu{\`e}re} C.-A.,  2021, \mn@doi [\mnras] {10.1093/mnras/stab1264}, \href {https://ui.adsabs.harvard.edu/abs/2021MNRAS.505..259J} {505, 259}

\bibitem[\protect\citeauthoryear{{Jiang} \& {Oh}}{{Jiang} \& {Oh}}{2018}]{Jiang2018}
{Jiang} Y.-F.,  {Oh} S.~P.,  2018, \mn@doi [\apj] {10.3847/1538-4357/aaa6ce}, \href {https://ui.adsabs.harvard.edu/abs/2018ApJ...854....5J} {854, 5}

\bibitem[\protect\citeauthoryear{{Kempski} \& {Quataert}}{{Kempski} \& {Quataert}}{2022}]{Kempski2022}
{Kempski} P.,  {Quataert} E.,  2022, \mn@doi [\mnras] {10.1093/mnras/stac1240}, \href {https://ui.adsabs.harvard.edu/abs/2022MNRAS.514..657K} {514, 657}

\bibitem[\protect\citeauthoryear{{Kempski}, {Fielding}, {Quataert}, {Galishnikova}, {Kunz}, {Philippov}  \& {Ripperda}}{{Kempski} et~al.}{2023}]{Kempski2023}
{Kempski} P.,  {Fielding} D.~B.,  {Quataert} E.,  {Galishnikova} A.~K.,  {Kunz} M.~W.,  {Philippov} A.~A.,   {Ripperda} B.,  2023, \mn@doi [\mnras] {10.1093/mnras/stad2609}, \href {https://ui.adsabs.harvard.edu/abs/2023MNRAS.525.4985K} {525, 4985}

\bibitem[\protect\citeauthoryear{{Kormendy} \& {Ho}}{{Kormendy} \& {Ho}}{2013}]{Kormendy2013}
{Kormendy} J.,  {Ho} L.~C.,  2013, \mn@doi [\araa] {10.1146/annurev-astro-082708-101811}, \href {https://ui.adsabs.harvard.edu/abs/2013ARA&A..51..511K} {51, 511}

\bibitem[\protect\citeauthoryear{{Kravtsov}, {Vikhlinin}  \& {Meshcheryakov}}{{Kravtsov} et~al.}{2018}]{Kravtsov2018}
{Kravtsov} A.~V.,  {Vikhlinin} A.~A.,   {Meshcheryakov} A.~V.,  2018, \mn@doi [Astronomy Letters] {10.1134/S1063773717120015}, \href {https://ui.adsabs.harvard.edu/abs/2018AstL...44....8K} {44, 8}

\bibitem[\protect\citeauthoryear{{Kulsrud} \& {Pearce}}{{Kulsrud} \& {Pearce}}{1969}]{Kulsrud1969}
{Kulsrud} R.,  {Pearce} W.~P.,  1969, \mn@doi [\apj] {10.1086/149981}, \href {https://ui.adsabs.harvard.edu/abs/1969ApJ...156..445K} {156, 445}

\bibitem[\protect\citeauthoryear{{Lemoine}}{{Lemoine}}{2023}]{Lemoine2023}
{Lemoine} M.,  2023, \mn@doi [Journal of Plasma Physics] {10.1017/S0022377823000946}, \href {https://ui.adsabs.harvard.edu/abs/2023JPlPh..89e1701L} {89, 175890501}

\bibitem[\protect\citeauthoryear{{Lerche}}{{Lerche}}{1967}]{Lerche1967}
{Lerche} I.,  1967, \mn@doi [\apj] {10.1086/149045}, \href {https://ui.adsabs.harvard.edu/abs/1967ApJ...147..689L} {147, 689}

\bibitem[\protect\citeauthoryear{{Li} \& {Bryan}}{{Li} \& {Bryan}}{2014}]{Li2014}
{Li} Y.,  {Bryan} G.~L.,  2014, \mn@doi [\apj] {10.1088/0004-637X/789/1/54}, \href {https://ui.adsabs.harvard.edu/abs/2014ApJ...789...54L} {789, 54}

\bibitem[\protect\citeauthoryear{{Mao} \& {Ostriker}}{{Mao} \& {Ostriker}}{2018}]{Mao2018}
{Mao} S.~A.,  {Ostriker} E.~C.,  2018, \mn@doi [\apj] {10.3847/1538-4357/aaa88e}, \href {https://ui.adsabs.harvard.edu/abs/2018ApJ...854...89M} {854, 89}

\bibitem[\protect\citeauthoryear{{McCarthy} et~al.,}{{McCarthy} et~al.}{2024}]{McCarthy2024}
{McCarthy} I.~G.,  et~al., 2024, \mn@doi [arXiv e-prints] {10.48550/arXiv.2410.19905}, \href {https://ui.adsabs.harvard.edu/abs/2024arXiv241019905M} {p. arXiv:2410.19905}

\bibitem[\protect\citeauthoryear{{Moster}, {Naab}  \& {White}}{{Moster} et~al.}{2013}]{Moster2013}
{Moster} B.~P.,  {Naab} T.,   {White} S. D.~M.,  2013, \mn@doi [\mnras] {10.1093/mnras/sts261}, \href {https://ui.adsabs.harvard.edu/abs/2013MNRAS.428.3121M} {428, 3121}

\bibitem[\protect\citeauthoryear{{Narayan} \& {Medvedev}}{{Narayan} \& {Medvedev}}{2001}]{Narayan2001}
{Narayan} R.,  {Medvedev} M.~V.,  2001, \mn@doi [\apjl] {10.1086/338325}, \href {https://ui.adsabs.harvard.edu/abs/2001ApJ...562L.129N} {562, L129}

\bibitem[\protect\citeauthoryear{{Pandey} et~al.,}{{Pandey} et~al.}{2022}]{Sandey2022}
{Pandey} S.,  et~al., 2022, \mn@doi [\prd] {10.1103/PhysRevD.105.123526}, \href {https://ui.adsabs.harvard.edu/abs/2022PhRvD.105l3526P} {105, 123526}

\bibitem[\protect\citeauthoryear{{Pinzke} \& {Pfrommer}}{{Pinzke} \& {Pfrommer}}{2010}]{Pinzke2010}
{Pinzke} A.,  {Pfrommer} C.,  2010, \mn@doi [\mnras] {10.1111/j.1365-2966.2010.17328.x}, \href {https://ui.adsabs.harvard.edu/abs/2010MNRAS.409..449P} {409, 449}

\bibitem[\protect\citeauthoryear{{Pratt}, {Arnaud}, {Maughan}  \& {Melin}}{{Pratt} et~al.}{2022}]{Pratt2022}
{Pratt} G.~W.,  {Arnaud} M.,  {Maughan} B.~J.,   {Melin} J.~B.,  2022, \mn@doi [\aap] {10.1051/0004-6361/202243074}, \href {https://ui.adsabs.harvard.edu/abs/2022A&A...665A..24P} {665, A24}

\bibitem[\protect\citeauthoryear{{Preston}, {Amon}  \& {Efstathiou}}{{Preston} et~al.}{2023}]{Preston2023}
{Preston} C.,  {Amon} A.,   {Efstathiou} G.,  2023, \mn@doi [\mnras] {10.1093/mnras/stad2573}, \href {https://ui.adsabs.harvard.edu/abs/2023MNRAS.525.5554P} {525, 5554}

\bibitem[\protect\citeauthoryear{{Quataert}, {Jiang}  \& {Thompson}}{{Quataert} et~al.}{2022}]{Quataert2022}
{Quataert} E.,  {Jiang} Y.-F.,   {Thompson} T.~A.,  2022, \mn@doi [\mnras] {10.1093/mnras/stab3274}, \href {https://ui.adsabs.harvard.edu/abs/2022MNRAS.510..920Q} {510, 920}

\bibitem[\protect\citeauthoryear{{Ramesh}, {Nelson}  \& {Girichidis}}{{Ramesh} et~al.}{2024}]{Ramesh2024}
{Ramesh} R.,  {Nelson} D.,   {Girichidis} P.,  2024, \mn@doi [arXiv e-prints] {10.48550/arXiv.2409.18238}, \href {https://ui.adsabs.harvard.edu/abs/2024arXiv240918238R} {p. arXiv:2409.18238}

\bibitem[\protect\citeauthoryear{{Reichherzer}, {Bott}, {Ewart}, {Gregori}, {Kempski}, {Kunz}  \& {Schekochihin}}{{Reichherzer} et~al.}{2023}]{Reichherzer2025}
{Reichherzer} P.,  {Bott} A. F.~A.,  {Ewart} R.~J.,  {Gregori} G.,  {Kempski} P.,  {Kunz} M.~W.,   {Schekochihin} A.~A.,  2023, \mn@doi [arXiv e-prints] {10.48550/arXiv.2311.01497}, \href {https://ui.adsabs.harvard.edu/abs/2023arXiv231101497R} {p. arXiv:2311.01497}

\bibitem[\protect\citeauthoryear{{Ruszkowski} \& {Pfrommer}}{{Ruszkowski} \& {Pfrommer}}{2023}]{Ruszkowski2023}
{Ruszkowski} M.,  {Pfrommer} C.,  2023, \mn@doi [\aapr] {10.1007/s00159-023-00149-2}, \href {https://ui.adsabs.harvard.edu/abs/2023A&ARv..31....4R} {31, 4}

\bibitem[\protect\citeauthoryear{{Schlickeiser}}{{Schlickeiser}}{2002}]{Schlickeiser2002}
{Schlickeiser} R.,  2002, {Cosmic Ray Astrophysics}

\bibitem[\protect\citeauthoryear{{Shen}, {Hopkins}, {Faucher-Gigu{\`e}re}, {Alexander}, {Richards}, {Ross}  \& {Hickox}}{{Shen} et~al.}{2020}]{Shen_qlf2020}
{Shen} X.,  {Hopkins} P.~F.,  {Faucher-Gigu{\`e}re} C.-A.,  {Alexander} D.~M.,  {Richards} G.~T.,  {Ross} N.~P.,   {Hickox} R.~C.,  2020, \mn@doi [\mnras] {10.1093/mnras/staa1381}, \href {https://ui.adsabs.harvard.edu/abs/2020MNRAS.495.3252S} {495, 3252}

\bibitem[\protect\citeauthoryear{{Simet}, {Battaglia}, {Mandelbaum}  \& {Seljak}}{{Simet} et~al.}{2017}]{Simet2017}
{Simet} M.,  {Battaglia} N.,  {Mandelbaum} R.,   {Seljak} U.,  2017, \mn@doi [\mnras] {10.1093/mnras/stw3322}, \href {https://ui.adsabs.harvard.edu/abs/2017MNRAS.466.3663S} {466, 3663}

\bibitem[\protect\citeauthoryear{{Skilling}}{{Skilling}}{1971}]{Skilling1971}
{Skilling} J.,  1971, \mn@doi [\apj] {10.1086/151210}, \href {https://ui.adsabs.harvard.edu/abs/1971ApJ...170..265S} {170, 265}

\bibitem[\protect\citeauthoryear{{Tchekhovskoy}, {Narayan}  \& {McKinney}}{{Tchekhovskoy} et~al.}{2011}]{Tchekhovskoy2011}
{Tchekhovskoy} A.,  {Narayan} R.,   {McKinney} J.~C.,  2011, \mn@doi [\mnras] {10.1111/j.1745-3933.2011.01147.x}, \href {https://ui.adsabs.harvard.edu/abs/2011MNRAS.418L..79T} {418, L79}

\bibitem[\protect\citeauthoryear{{Thomas} \& {Pfrommer}}{{Thomas} \& {Pfrommer}}{2019}]{Thomas2019}
{Thomas} T.,  {Pfrommer} C.,  2019, \mn@doi [\mnras] {10.1093/mnras/stz263}, \href {https://ui.adsabs.harvard.edu/abs/2019MNRAS.485.2977T} {485, 2977}

\bibitem[\protect\citeauthoryear{{Thompson} \& {Heckman}}{{Thompson} \& {Heckman}}{2024}]{Thompson2024}
{Thompson} T.~A.,  {Heckman} T.~M.,  2024, \mn@doi [\araa] {10.1146/annurev-astro-041224-011924}, \href {https://ui.adsabs.harvard.edu/abs/2024ARA&A..62..529T} {62, 529}

\bibitem[\protect\citeauthoryear{{Tompkins}, {Driver}, {Robotham}, {Windhorst}, {Lagos}, {Vernstrom}  \& {Hopkins}}{{Tompkins} et~al.}{2023}]{Tompkins2023}
{Tompkins} S.~A.,  {Driver} S.~P.,  {Robotham} A. S.~G.,  {Windhorst} R.~A.,  {Lagos} C. d.~P.,  {Vernstrom} T.,   {Hopkins} A.~M.,  2023, \mn@doi [\mnras] {10.1093/mnras/stad116}, \href {https://ui.adsabs.harvard.edu/abs/2023MNRAS.521..332T} {521, 332}

\bibitem[\protect\citeauthoryear{{Trotta}, {J{\'o}hannesson}, {Moskalenko}, {Porter}, {Ruiz de Austri}  \& {Strong}}{{Trotta} et~al.}{2011}]{Trotta2011}
{Trotta} R.,  {J{\'o}hannesson} G.,  {Moskalenko} I.~V.,  {Porter} T.~A.,  {Ruiz de Austri} R.,   {Strong} A.~W.,  2011, \mn@doi [\apj] {10.1088/0004-637X/729/2/106}, \href {https://ui.adsabs.harvard.edu/abs/2011ApJ...729..106T} {729, 106}

\bibitem[\protect\citeauthoryear{{Wellons} et~al.,}{{Wellons} et~al.}{2023}]{Wellons2023}
{Wellons} S.,  et~al., 2023, \mn@doi [\mnras] {10.1093/mnras/stad511}, \href {https://ui.adsabs.harvard.edu/abs/2023MNRAS.520.5394W} {520, 5394}

\bibitem[\protect\citeauthoryear{{Wentzel}}{{Wentzel}}{1971}]{Wentzel1971}
{Wentzel} D.~G.,  1971, \mn@doi [\apj] {10.1086/150794}, \href {https://ui.adsabs.harvard.edu/abs/1971ApJ...163..503W} {163, 503}

\bibitem[\protect\citeauthoryear{{Wiener}, {Oh}  \& {Guo}}{{Wiener} et~al.}{2013}]{Wiener2013}
{Wiener} J.,  {Oh} S.~P.,   {Guo} F.,  2013, \mn@doi [\mnras] {10.1093/mnras/stt1163}, \href {https://ui.adsabs.harvard.edu/abs/2013MNRAS.434.2209W} {434, 2209}

\bibitem[\protect\citeauthoryear{{Wiener}, {Pfrommer}  \& {Oh}}{{Wiener} et~al.}{2017}]{Wiener2017}
{Wiener} J.,  {Pfrommer} C.,   {Oh} S.~P.,  2017, \mn@doi [\mnras] {10.1093/mnras/stx127}, \href {https://ui.adsabs.harvard.edu/abs/2017MNRAS.467..906W} {467, 906}

\bibitem[\protect\citeauthoryear{{Wiener}, {Zweibel}  \& {Oh}}{{Wiener} et~al.}{2018}]{Wiener2018}
{Wiener} J.,  {Zweibel} E.~G.,   {Oh} S.~P.,  2018, \mn@doi [\mnras] {10.1093/mnras/stx2603}, \href {https://ui.adsabs.harvard.edu/abs/2018MNRAS.473.3095W} {473, 3095}

\bibitem[\protect\citeauthoryear{{Yan} \& {Lazarian}}{{Yan} \& {Lazarian}}{2002}]{Yan2002}
{Yan} H.,  {Lazarian} A.,  2002, \mn@doi [PRL] {10.1103/PhysRevLett.89.281102}, \href {https://ui.adsabs.harvard.edu/abs/2002PhRvL..89B1102Y} {89, 281102}

\bibitem[\protect\citeauthoryear{{Zhang}, {Finoguenov}, {B{\"o}hringer}, {Kneib}, {Smith}, {Kneissl}, {Okabe}  \& {Dahle}}{{Zhang} et~al.}{2008}]{Zhang2008}
{Zhang} Y.~Y.,  {Finoguenov} A.,  {B{\"o}hringer} H.,  {Kneib} J.~P.,  {Smith} G.~P.,  {Kneissl} R.,  {Okabe} N.,   {Dahle} H.,  2008, \mn@doi [\aap] {10.1051/0004-6361:20079103}, \href {https://ui.adsabs.harvard.edu/abs/2008A&A...482..451Z} {482, 451}

\bibitem[\protect\citeauthoryear{{Zweibel}}{{Zweibel}}{2017}]{Zweibel2017}
{Zweibel} E.~G.,  2017, \mn@doi [Physics of Plasmas] {10.1063/1.4984017}, \href {https://ui.adsabs.harvard.edu/abs/2017PhPl...24e5402Z} {24, 055402}

\bibitem[\protect\citeauthoryear{{van Daalen}, {McCarthy}  \& {Schaye}}{{van Daalen} et~al.}{2020}]{vanDaalen2020}
{van Daalen} M.~P.,  {McCarthy} I.~G.,   {Schaye} J.,  2020, \mn@doi [\mnras] {10.1093/mnras/stz3199}, \href {https://ui.adsabs.harvard.edu/abs/2020MNRAS.491.2424V} {491, 2424}

\bibitem[\protect\citeauthoryear{{van Weeren}, {de Gasperin}, {Akamatsu}, {Br{\"u}ggen}, {Feretti}, {Kang}, {Stroe}  \& {Zandanel}}{{van Weeren} et~al.}{2019}]{vanWeeren2019}
{van Weeren} R.~J.,  {de Gasperin} F.,  {Akamatsu} H.,  {Br{\"u}ggen} M.,  {Feretti} L.,  {Kang} H.,  {Stroe} A.,   {Zandanel} F.,  2019, \mn@doi [\ssr] {10.1007/s11214-019-0584-z}, \href {https://ui.adsabs.harvard.edu/abs/2019SSRv..215...16V} {215, 16}

\makeatother
\end{thebibliography}


\end{document}